\documentclass[prd,aps,nofootinbib,twocolumn,preprintnumbers]{revtex4}
\usepackage[utf8]{inputenc}
\usepackage[colorlinks=true,citecolor=blue,linkcolor=blue]{hyperref}
\usepackage[normalem]{ulem}
\usepackage{amsmath,amssymb, mathrsfs}
\usepackage{epsfig}
\usepackage{graphicx}               
\usepackage{url}
\usepackage{color}
\usepackage{slashed}
\usepackage{multirow}
\usepackage{placeins}
\usepackage[dvipsnames]{xcolor}
\usepackage{epstopdf}
\usepackage{soul}
\usepackage{tikz}
\usepackage[capitalise, english]{cleveref}
\usepackage{siunitx}
\usepackage{xspace}
\usepackage{booktabs}
\usepackage{verbatim}
\usetikzlibrary{trees}
\usetikzlibrary{decorations.pathmorphing}
\usetikzlibrary{decorations.markings}

\newcommand\myshade{80}
\colorlet{mylinkcolor}{ForestGreen}
\colorlet{mycitecolor}{Red}
\colorlet{myurlcolor}{violet}

\hypersetup{
	linkcolor  = mylinkcolor!\myshade!black,
	citecolor  = mycitecolor!\myshade!black,
	urlcolor   = myurlcolor!\myshade!black,
	colorlinks = true
}

\definecolor{jblue}{RGB}{20,50,100}
\definecolor{npurple}{RGB} {153, 51, 204}
\definecolor{wred}{RGB}{217,0,56}
\definecolor{white}{RGB}{255,255,255}

\definecolor{korange}{RGB}{235, 80,  43}
\definecolor{korange2}{RGB}{245, 100,  63}
\definecolor{kyelloworange}{RGB}{255, 210,  110}
\definecolor{kyelloworange2}{RGB}{240, 170,  90}
\definecolor{kred}{RGB}{204,  102, 153}
\definecolor{kpurple}{RGB}{153,  61, 190}
\definecolor{kpurplelight}{RGB}{213,  161, 230}

\definecolor{tobycolour}{rgb}{.5,.0,.5}

\DeclareSIUnit\year{yr}
\DeclareSIUnit\pc{pc}
\DeclareSIUnit\ergs{ergs}
\DeclareSIUnit\msun{\ensuremath{M_\odot}}
\allowdisplaybreaks

\makeatletter
\providecommand*{\diff}%
{\@ifnextchar^{\DIfF}{\DIfF^{}}}
\def\DIfF^#1{%
	\mathop{\mathrm{\mathstrut d}}%
	\nolimits^{#1}\gobblespace}
\def\gobblespace{%
	\futurelet\diffarg\opspace}
\def\opspace{%
	\let\DiffSpace\!%
	\ifx\diffarg(%
	\let\DiffSpace\relax
	\else
	\ifx\diffarg[%
	\let\DiffSpace\relax
	\else
	\ifx\diffarg\{%
	\let\DiffSpace\relax
	\fi\fi\fi\DiffSpace}

\usepackage{tikz,xcolor,hyperref}

\definecolor{lime}{HTML}{A6CE39}
\DeclareRobustCommand{\orcidicon}{\hspace{-1mm}
	\begin{tikzpicture}
		\draw[lime, fill=lime] (0,0) 
		circle [radius=0.16] 
		node[white] {{\fontfamily{qag}\selectfont \tiny \,ID}};
		\draw[white, fill=white] (-0.0525,0.095) 
		circle [radius=0.007];
	\end{tikzpicture}
	\hspace{-3mm}
}

\foreach \x in {A, ..., Z}{\expandafter\xdef\csname orcid\x\endcsname{\noexpand\href{https://orcid.org/\csname orcidauthor\x\endcsname}
		{\noexpand\orcidicon}}
}

\keywords{}

\begin{document}
	
	\title{Sensitivities on non-spinning and spinning primordial black hole dark matter with global 21 cm troughs}
	
\author{Akash Kumar Saha\orcidB{}} 
\email{akashks@iisc.ac.in}
\affiliation{Centre for High Energy Physics, Indian Institute of Science, C.\,V.\,Raman Avenue, Bengaluru 560012, India}

\author{Ranjan Laha\orcidD{}} 
\email{ranjanlaha@iisc.ac.in}
\affiliation{Centre for High Energy Physics, Indian Institute of Science, C.\,V.\,Raman Avenue, Bengaluru 560012, India}

	\date{\today}
	
	
	\begin{abstract}
	 Detection of the global 21 cm signal arising from neutral hydrogen can revolutionize our understanding of the standard evolution of the universe after recombination. In addition, it can also be an excellent probe of Dark Matter (DM). Among all the DM candidates, Primordial Black Holes (PBHs) are one of the most well-motivated. Hawking emission from low-mass PBHs can have substantial effect on the thermal and ionization history of the early universe, and that in turn can have an imprint on the global 21 cm signal. Recently EDGES has claimed a global 21 cm signal, though SARAS 3 has rejected that claim. In this work, we investigate the sensitivities on non-spinning and spinning PBHs arising from an EDGES-like measurement of the global 21 cm signal, and find that the sensitivities will be competitive with those arising from other astrophysical observables. We show that the sensitivities can be significantly strengthened depending on various uncertain astrophysical parameters.  Besides, we also derive projections on the PBH density from the absorption trough expected during the Dark Ages. Our work shows that the near future unambiguous  detection of the global 21 cm absorption troughs can be an excellent probe of PBH DM.   
	\end{abstract}
	
	\maketitle
	
	\section{Introduction}
	\label{sec:introduction}
The prospect of detecting global 21 cm signal has been of interest for many years\,\cite{Furlanetto:2019jso,Furlanetto:2019jzo,Liu:2019srd,pritchard201221,Padmanabhan:2021hkg}. The early universe, after recombination, was mostly filled with neutral hydrogen. The redshifted 21 cm line arising from the hyperfine splitting of the ground state of neutral hydrogen\,\cite{pritchard201221} can work as an important tracer to the temperature and ionization history of the Intergalactic Medium (IGM). From the astrophysical point of view, this can shed light on the exact epoch of reionization and the formation of earliest stars and galaxies \,\cite{Mirocha:2018cih,Park:2018ljd,Munoz:2019hjh,Munoz:2021psm,Magg:2021jyc}. It can also provide important insights into the cosmology and fundamental physics of the early universe, esp.\,\,of various beyond the Standard Model scenarios. Recently, the first possible and controversial observation of global 21 cm signal by the EDGES collaboration \cite{Bowman:2018yin} has rejuvenated interest in this field. The collaboration has detected an absorption trough with differential brightness temperature, $T_{21}= -500^{+200}_{-500}$\, mK at 78 MHz ($z\sim17.2$) with 99\% confidence level. The amplitude of this measured absorption trough is more than two times larger
	than the theoretical prediction from the $\Lambda$ cold dark matter ($\Lambda$CDM)  model of the universe\,\cite{PhysRevD.83.043513,10.1111/j.1365-2966.2010.17940.x,pritchard201221,Fialkov:2019vnb,Xu:2021zkf}. This has resulted in plethora of research on the origin of the anomaly \,\cite{Fraser:2018acy,Liu:2018uzy,Barkana:2018qrx,Feng:2018rje,Ewall-Wice:2018bzf,Munoz:2018pzp,Fialkov:2018xre,McGaugh:2018ysb,Berlin:2018sjs,Yang:2018gjd,Pospelov:2018kdh,Costa:2018aoy,Falkowski:2018qdj,Lambiase:2018lhs,Sharma:2018agu,Jia:2018csj,Moroi:2018vci,Houston:2018vrf,Sikivie:2018tml,Auriol:2018ovo,Li:2018okf,Dhuria:2018kzb,Munoz:2018jwq,Bhatt:2019qbq,Liu:2019knx,Choi:2019jwx,Natwariya:2020mhe,Johns:2020rtp,Mathur:2021gej,Melia:2021anz, Dhuria:2021lqs,Theriault:2021mrq,Aboubrahim:2021ohe,Li:2021zcy}. The EDGES result has also stirred up a lot of controversy\,\cite{Hills:2018vyr,2018Natur.564E..35B,Bradley:2018eev,Tauscher:2020wso,Singh:2019gsv}. More recently, SARAS 3 has performed an independent check on this signal. SARAS 3 rejected EDGES signal with 95.3\% confidence level\,\cite{Singh:2021mxo}. In spite of this very recent development, we use the EDGES measurement in this paper as a representative of any global 21 cm excess that may potentially be observed in the near future.

The matter content of our present universe is dominated by DM. The true nature of DM has been a mystery for a long time. Although we have detected the gravitational signatures of DM over a variety of length scales, we are yet to detect any other fundamental interaction of it\,\cite{Green:2021jrr,Strigari:2012acq,Arbey:2021gdg,Brems:2018gqx,Lisanti:2016jxe}. Of all the DM candidates, PBHs are one of the oldest (see ref.\,\cite{Bertone:2016nfn} for a historical review of DM and refs.\,\cite{Green:2020jor,Carr:2021bzv,Carr:2020xqk,Carr:2020gox,Villanueva-Domingo:2021spv,Byrnes:2021jka} for recent reviews on PBH DM). PBHs are formed from large density perturbations or from other exotic mechanisms in the early universe (see for e.g., \cite{Bhaumik:2019tvl,Cai:2019bmk,Bhaumik:2020dor,Khlopov:2008qy,Belotsky:2018wph,sym11040511}. Depending on their formation time\,\cite{Carr:2020xqk}, the mass of a PBH is thought to lie anywhere between Planck mass and few hundred solar masses. A PBH both Hawking radiates \,\cite{Hawking:1975vcx} and accretes, although Hawking emission dominates for low-mass PBHs (in this work, we will refer to PBHs in mass range $10^{15}-10^{18}$\,g as low-mass PBHs). Accretion dominates for BHs with masses greater than a few solar mass\,\cite{DeLuca:2020fpg}. Non-spinning PBHs with masses $\lesssim 6\times10^{14}$\,g have evaporated by now\,\cite{Page:1976df,Arbey:2019jmj}. Extensive work has been done regarding the search for evaporating PBHs as DM candidate from various astrophysical observables\,\cite{Laha:2020ivk,1991ApJ...371..447M,Poulin_2017,Laha:2019ssq,PhysRevLett.122.041104,PhysRevLett.126.171101,PhysRevLett.125.101101,PhysRevD.101.023010,Laha:2020vhg,PhysRevD.95.083006,Lee:2021qhe,Chan:2020zry,Poulin:2016anj,Siegert:2021upf,Iguaz:2021irx,Keith:2021guq,Domenech:2021wkk,Schiavone:2021imu,Cai:2020fnq,Belotsky:2014kca,Ballesteros:2019exr}.

A low-mass PBH emits all sorts of particles, and these can subsequently heat up the IGM depending on particle interactions. As the amplitude of the absorption trough in a global 21 cm signal depends on the temperature and ionization of the IGM at a given redshift, if low-mass PBH DM exists, they can affect the observed global 21 cm signal. 

Various authors have already considered similar ideas \,\cite{Clark:2018ghm,Halder:2021rrh,Halder:2021rbq,Halder:2021jiv,Yang:2020egn,Yang:2020zcu,Mittal:2021egv,Natwariya:2021xki,Cang:2021owu}. In this work, in order to constrain PBH DM, we consider the absorption troughs at redshift $\sim$17 (EDGES-like measurement) and the one expected at Dark Ages. The latter trough solely depends on the underlying cosmology and not on various astrophysical uncertainties. For the EDGES-like measurement, we consider excess background radiation and non-standard recombination as explanations of the signal. In addition, we also consider the effects a power-law excess background radiation (that is consistent with EDGES-like signal) on the Dark Ages absorption trough. 

We structure this paper as follows. In section (\ref{sec:21 cm Cosmology}) we briefly review the underlying physics of 21 cm cosmology and the origin of both the absorption troughs. In section (\ref{standard thermal history}) we discuss the standard thermal history of IGM in the absence of any exotic physics. In section (\ref{PBH review}) PBH evaporation is discussed in detail. In section (\ref{results}) we display our results, and we conclude in section (\ref{conclusion}).

\section{21 cm Cosmology}
\label{sec:21 cm Cosmology}
21 cm line is an important tool in modern astrophysics and cosmology. Due to the interaction of magnetic moments of electron and proton in neutral hydrogen, we have a splitting in the ground state energy level. The energy difference between the levels corresponds to a photon of frequency 1420 MHz or wavelength of around 21 cm. The global 21 cm signal is represented by the differential brightness temperature\,\cite{Zaldarriaga_2004}
\begin{eqnarray}
	\label{Brightness temp}
	\delta T_b=T_{21}\approx x_{\rm HI}(z)\left(\frac{0.15}{\Omega_m}\right)^{1/2} \left(\frac{\Omega_bh}{0.02}\right)\nonumber\\
	\times\left(\frac{1+z}{10}\right)^{1/2}\left(1-\frac{T_R(z)}{T_S(z)}\right)23 \,\,\text{mK},
\end{eqnarray} 
where $x_{\rm HI} (z)$ is the neutral hydrogen fraction as a function of redshift $z$, $\Omega_m$ and $\Omega_b$ are the matter and baryon energy density parameter today respectively, $h$ is the current Hubble parameter in units of 100 km s$^{-1}$Mpc$^{-1}$, $T_R$ is the background radiation temperature, and $T_S$ is the spin temperature which is also known as 21 cm excitation temperature. $T_S$ is defined as the ratio between the number densities ($n_i$) of hydrogen atoms in the two hyperfine levels
\begin{eqnarray}
	\frac{n_1}{n_0}=\frac{g_1}{g_0}e^{-T_*/T_S},
\end{eqnarray}
where subscript 0 and 1 are for 1S singlet and triplet states respectively. The ratio of statistical degeneracy factors of two levels is denoted by $g_1/g_0$ and $T_*=hc/k \lambda_{21 cm}=0.068$ K\,\cite{pritchard201221}.
Spin temperature can also be written as
\begin{eqnarray}
	T_S^{-1}=\frac{T_R^{-1}+x_\alpha T_\alpha^{-1}+x_cT_K^{-1}}{1+x_\alpha+x_c},
\end{eqnarray}
where $T_\alpha$ and $T_K$ are the colour temperature of Lyman-$\alpha$ radiation field and gas kinetic temperature, respectively. The terms $x_\alpha$, $x_c$ are the coupling coefficients due to Lyman-$\alpha$ photon scattering and atomic collisions respectively. \\
Eq.\,(\ref{Brightness temp}) implies that an absorption signal ($\delta T_b<0$) requires $T_S<T_R$. Below we will briefly discuss the origin of two absorption troughs expected in the global 21 cm signal\,\cite{pritchard201221}.
\subsection{Dark Ages Trough}
\label{DarkAgesTrough}
 After recombination, CMB temperature was coupled to the gas temperature ($T_m$) through Compton scattering with the residual free electrons, implying $T_m\approx T_R$ (assuming $T_R=T_{\rm CMB}$). On the other hand, collisions between neutral hydrogen atoms, free protons and free electrons in the gas induced 21 cm transitions, which ensures $T_S\approx T_m$. The above two conditions imply $T_S\approx T_R$, which in turn results in $\delta T_b\approx 0$ from eq.\,\eqref{Brightness temp}. Thus, we cannot have a 21 cm signal from this period. This Compton scattering with residual electrons is only efficient until $(1+z) \sim 155$. Afterwards due to adiabatic cooling, the gas temperature evolves as, $T_m\propto(1+z)^2$, thus falling faster than the CMB temperature, $T_{\rm CMB}\propto(1+z)$. This ensures that $T_m<T_R$. Collisional coupling still prevailed and as a result we get $T_S<T_R$ during this period. This results in $\delta T_b<0$ and we have the first absorption trough at the Dark Ages ($35\lesssim z\lesssim200$). The absorption trough ends at the point where collisional coupling becomes ineffective (due to low density of gas) and radiative coupling with CMB ensures $T_S=T_R$.
\subsection{Pre-reionization Trough} As the first stars begin to form, the Lyman-$\alpha$ photons from them heat up the gas and create 21 cm transitions. Lyman-$\alpha$ photons can excite an atom in the ground state to any one of the excited states (depending on the selection rule) and the atom can later de-excite via emission of a Lyman-$\alpha$ photon. During de-excitation, the atom can return to either of the ground state hyperfine levels. If an atom initially at hyperfine singlet state (triplet state), returns later to the triplet state (singlet state), spin flip transition occurs. Due to this Lyman-$\alpha$ coupling (Wouthuysen-Field effect\,\cite{1952Phy....18...75W,1958PIRE...46..240F}), we have $T_m\leq T_S<T_R$. This results in the second absorption trough. The depth of this trough depends on various uncertain astrophysical parameters\,\cite{Furlanetto:2006jb,Mittal:2020kjs}. EDGES has recently claimed a  detection of this trough with $T_{21}= -500^{+200}_{-500}$\, mK at redshift $z\sim17.2$ with 99\% confidence level, although SARAS 3 has rejected the claim at 95.3\% confidence level. The end of this trough is marked by the saturation of the effect of Lyman-$\alpha$ coupling ($x_{\alpha}>>1$) and soon heating becomes effective. Due to heating once we have $T_m\approx T_R$, we do not have any further 21 cm absorption signal.
\section{Standard Thermal and Ionization History}
\label{standard thermal history}
Before proceeding to the exotic heating due to PBHs, we first briefly review the standard thermal and ionization history. This is well described by TLA (Three-level atom) model which simplifies the infinite energy levels of hydrogen atom into three levels\,--- ground state ($n=0$), first excited state ($n=1$), and continuum state ($n=\infty$)\,\,\cite{1968ApJ...153....1P,1969JETP...28..146Z}. The temperature and ionization evolution can be described the following differential equations \cite{Liu:2018uzy} 
\begin{eqnarray}
	\label{tm}
	\frac{dT_m^{(0)}}{dt}=-2HT_m^{(0)}+\Gamma_c(T_{\rm CMB}-T_m^{(0)}),
\end{eqnarray}
\begin{eqnarray}
	\label{xe}
	\frac{dx_e^{(0)}}{dt}=-\mathcal{C}\left[n_H{(x_e^{(0)})}^2\alpha_B -4(1-x_e^{(0)})\beta_Be^{\frac{-E_{21}}{T_{\rm CMB}}}\right].
\end{eqnarray}
Here $T_m^{(0)}$ is the standard gas temperature, $t$ denotes time, $H$ is the Hubble parameter, $\Gamma_c$ is the Compton scattering rate, $T_{\rm CMB}$ is the CMB temperature, $x_e^{(0)}=n_{e}/n_H$, where $n_e$, $n_H$ are the number densities of free electrons and hydrogen atoms respectively. Peebles-C factor is denoted by $\mathcal{C}$, $\alpha_B$,\,$\beta_B$ are the case-B recombination and photoionization coefficients and $E_{21}=10.2$ eV is the Lyman-$\alpha$ transition energy. The Compton scattering rate ($\Gamma_c$) is given by
\begin{eqnarray}
	\Gamma_c=\frac{x_e^{(0)}}{1+f_{He}+x_e^{(0)}}\frac{8\sigma_Ta_rT_{\rm CMB}^4}{3m_ec},
\end{eqnarray}
	where $f_{He}=n_{He}/n_H$ is the relative abundance of helium nuclei by number, $\sigma_T$ is the Thomson cross section,
	$a_r$ is the radiation constant, and $m_e$ is the electron mass. In the presence of exotic heating due to new physics, both eqns.\,\eqref{tm} and \eqref{xe} get modified.
\section{Evaporation of Non-Spinning and Spinning PBH}
\label{PBH review}
\begin{figure}
	\centering
	\includegraphics[width=\columnwidth]{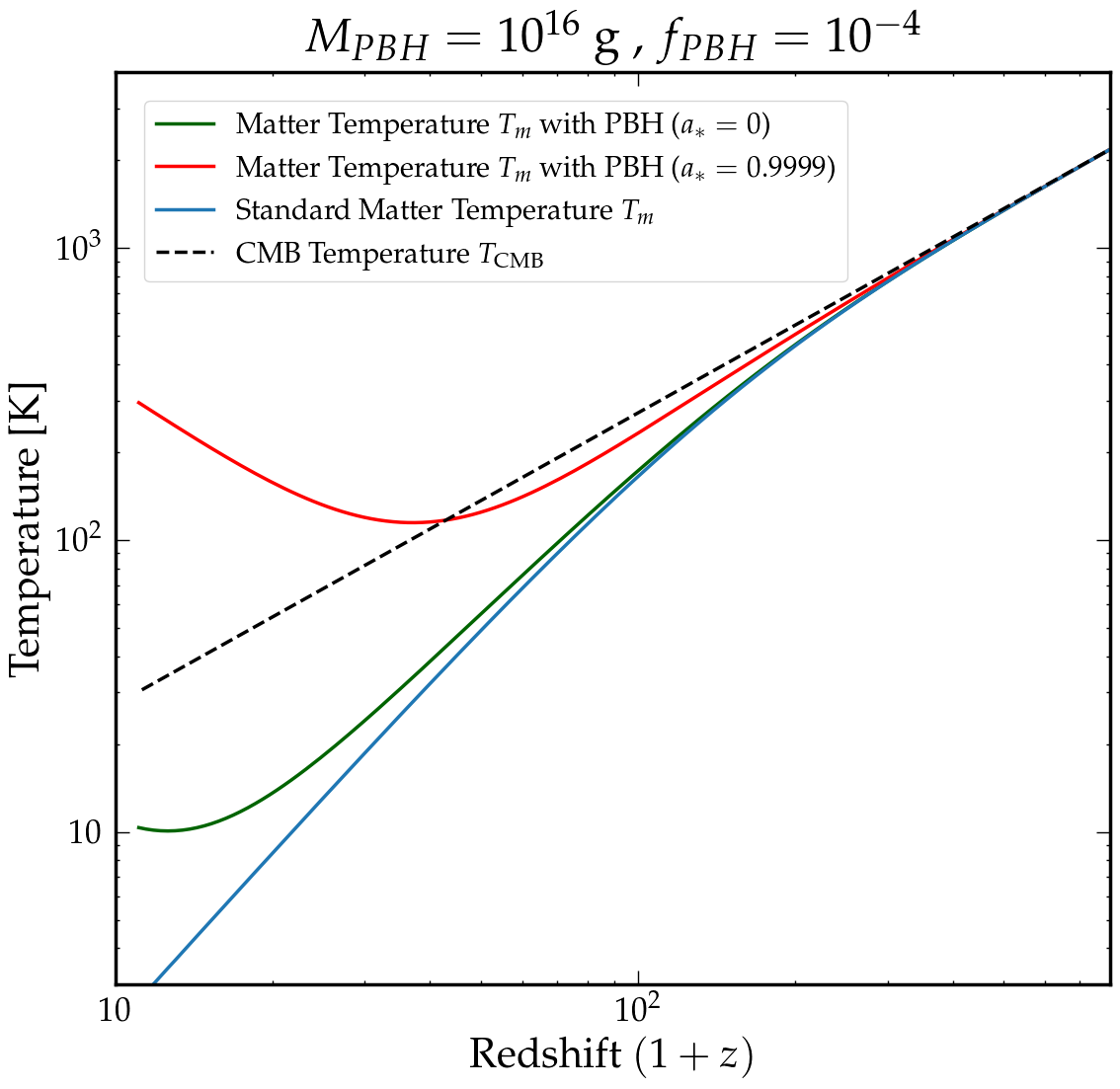}
	\caption{Thermal history of IGM in the absence and presence of PBH DM with masses = $10^{16}$\,g (monochromatic distribution) and $f_{PBH}$ = $10^{-4}$. Here the temperature evolution of CMB (black dotted line) and standard matter temperature evolution (blue line) are shown in comparison with matter temperature evolution in the presence of non-spinning (green line) and spinning (red line) PBH DM .} 
	\label{fig:Tm} 
\end{figure}
\label{sec:Evaporation of Non-Spinning and Spinning PBH}

The temperature of an uncharged, spinning PBH is given by\,\cite{Page:1976df,Page:1976ki,MacGibbon:1990zk}
\begin{eqnarray}
	T_{\rm PBH}=\frac{\hbar c^3}{4\pi G M_{\rm PBH}}\left(\frac{\sqrt{1-a_*^2}}{1+\sqrt{1-a_*^2}}\right),
	\label{eqn:1}
\end{eqnarray}
	where, $\hbar$ is the reduced Planck's constant, $c$ is the speed of light,  $G$ is the gravitational constant, $M_{\rm PBH}$ is the mass of the PBH and 
\begin{eqnarray}
	a_*=\frac{c^4J}{G M_{\rm PBH}^2},
\end{eqnarray}
	with $a_*$ being the the reduced
	spin parameter, and $J$ is the magnitude of angular momentum of the
	PBH. 
For $a_*\rightarrow0$, from eq.\,(\ref{eqn:1}) we get the temperature of the uncharged, non-spinning PBH\,\cite{MacGibbon:1990zk} as
\begin{eqnarray}
	T_{\rm PBH}=\frac{\hbar c^3}{8\pi G M_{\rm PBH}}=1.06\left(\frac{10^{13} \,\text{g}}{M_{\rm PBH}}\right) \text{GeV}.
\end{eqnarray}	
 \begin{figure*}[!htbp]
	\begin{center}
		\includegraphics[height=8cm]{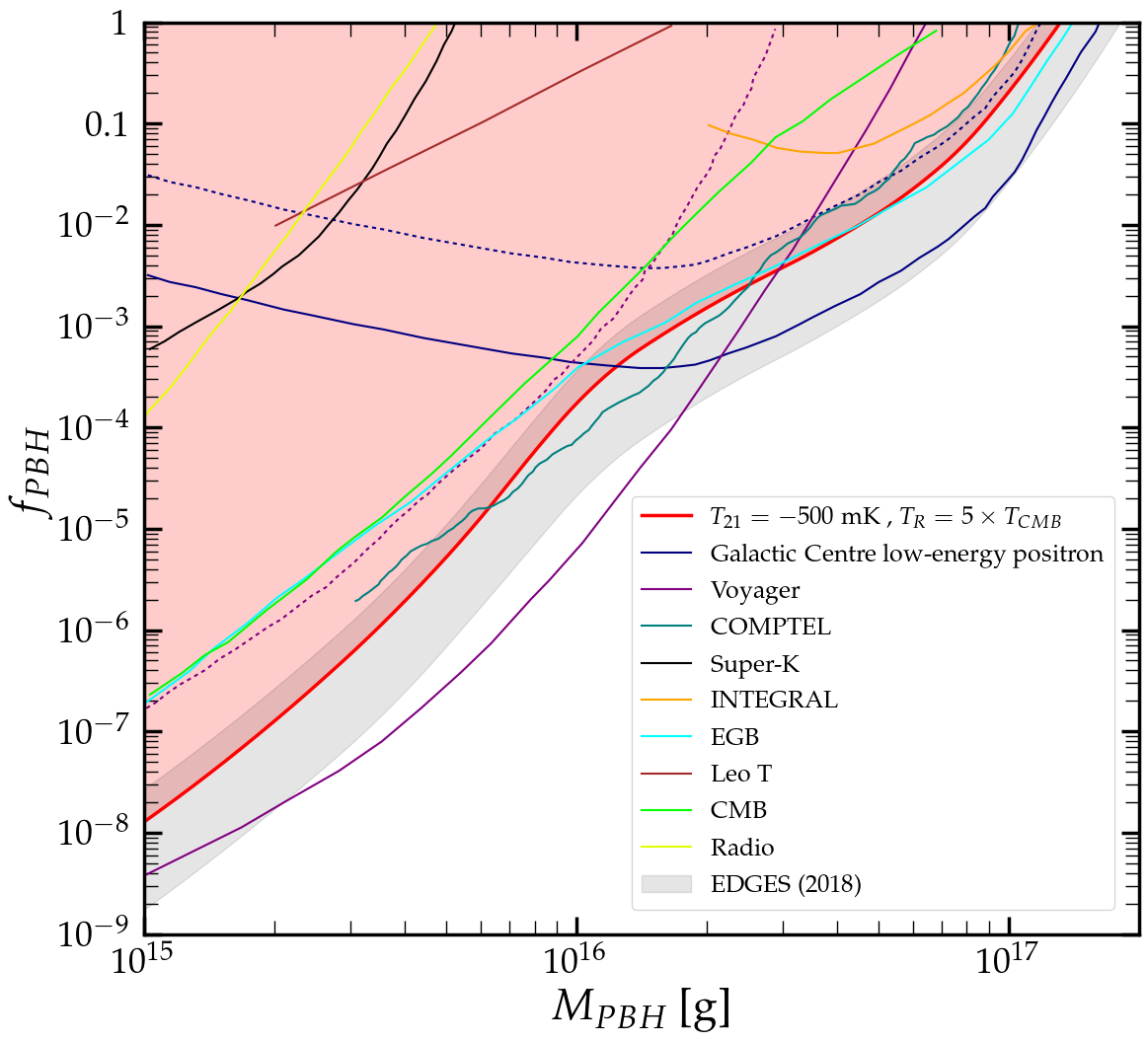}~~
		\includegraphics[height=8cm]{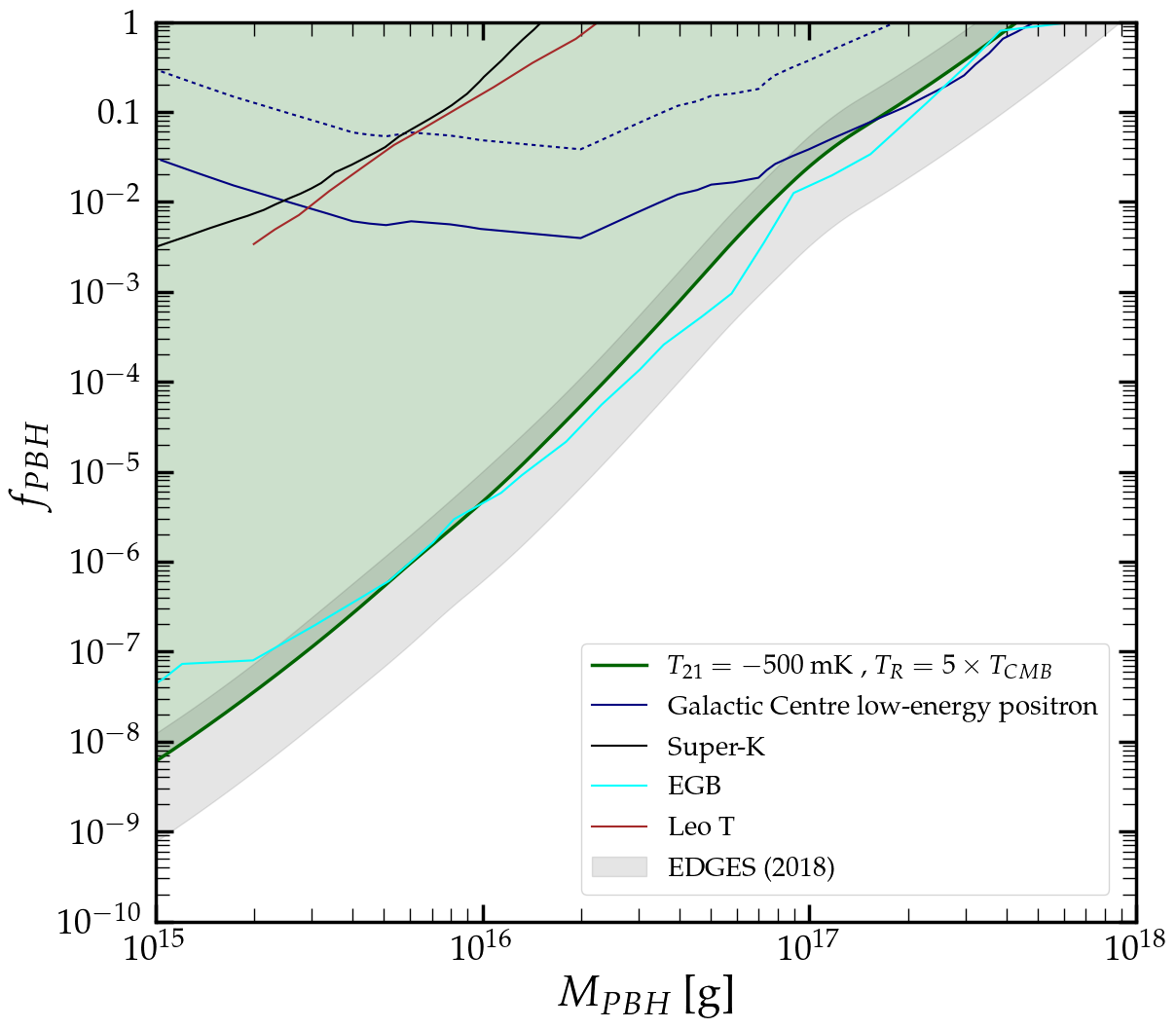}~~\\	
		\caption{Sensitivities on non-spinning ($a_*$=0) and spinning ($a_*$=0.9999) PBH DM (monochromatic distribution) from an assumed 21 cm absorption trough, similar to the one detected by EDGES  ($T_{21}=-500^{+200}_{-500}$\, mK), shown by the red (left panel) and green (right panel) lines respectively. For these limits, we assume $T_R=5\,T_{\rm CMB}$ ($z=17.2$). The grey shaded regions show the impact of the EDGES measurement uncertainty. Previous constraints include  Galactic Centre low-energy positron measurements by INTEGRAL (navy dotted line for  isothermal with 1.5 kpc and navy continuous line for NFW with 3.5 kpc region of interest)\,\cite{Laha:2019ssq}, measurement of the cosmic-ray flux by Voyager (purple dotted line for propagation model B without background and purple continuous line for propagation model A with background)\,\cite{PhysRevLett.122.041104}, COMPTEL (teal) and INTEGRAL (orange) measurements of the Galactic-centre keV-MeV gamma-ray flux \,\cite{PhysRevLett.126.171101,Laha:2020ivk}, Super-Kamiokande search for diffuse supernovae neutrino background (black)\,\cite{PhysRevLett.125.101101}, extra-Galactic gamma-ray emission measurement (aqua) \,\cite{PhysRevD.101.023010,Chen:2021ngo},  Leo T gas heating (brown)\,\cite{Laha:2020vhg} (see however ref.\,\cite{Kim:2020ngi} also), measurement of CMB from PLANCK (lime)\,\cite{PhysRevD.95.083006}, and radio measurement of the inner Galactic Centre (NFW profile) (lemon lime)\,\cite{Chan:2020zry}. Some of the constraints for non-spinning PBH DM are absent for spinning PBH DM because they are not present in the literature.}
		\label{fig: -ExcessTR}
	\end{center}	
\end{figure*}
\begin{figure}
	\centering
	\includegraphics[width=\columnwidth]{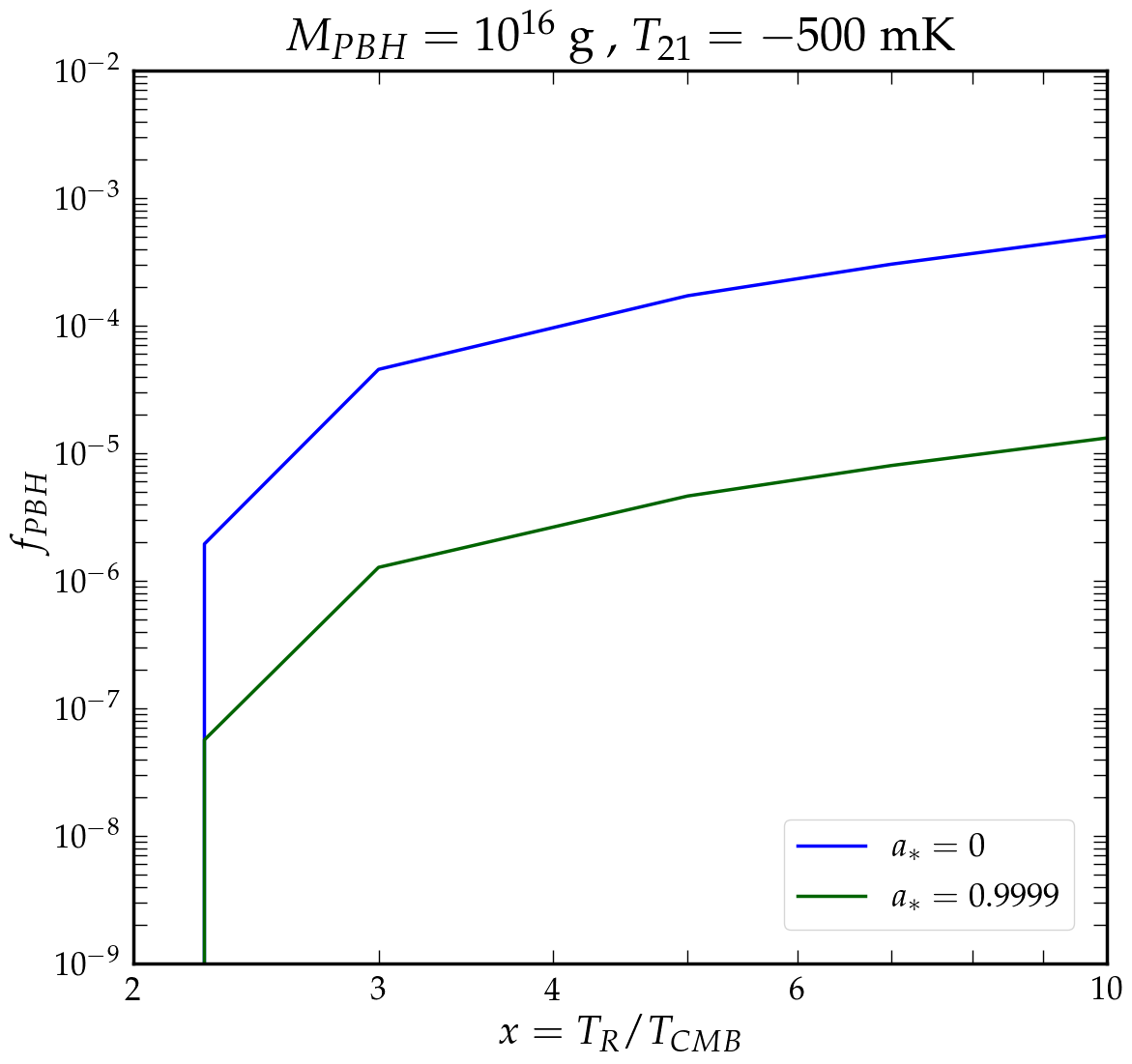}
	\caption{EDGES-like ($T_{21}=-500^{+200}_{-500}$\, mK) measurement sensitivities on non-spinning ($a_*$=0) and spinning ($a_*$=0.9999) PBH DM (monochromatic distribution) of mass 10$^{16}$\,g as a function of excess radiation given as integer multiples of $T_{\rm CMB}$ at redshift of interest ($z=17.2$).}
	\label{fig:-DiffTR} 
\end{figure}	
Throughout our analysis we assume uncharged PBHs. A charged low-mass PBH can discharge through pair production and it will have very little effect on the Hawking emission. Charge will only be an  important parameter for PBHs with masses $\gtrsim$ $10^5 M_{\odot}$\,\cite{Page:1976df,Arbey:2019mbc,Arbey:2021mbl}.

For primary emission, the number of particles, $N_i$ emitted per unit energy per unit time is given by\,\,\cite{Page:1976df,Page:1976ki,MacGibbon:1990zk}
\begin{eqnarray}
	\frac{d^2N_i}{dE dt}=\frac{1}{2 \pi\hbar}\sum_{dof}\frac{\Gamma_i(E,M_{\rm PBH},a_*)}{e^{E'/kT}\pm 1},
\end{eqnarray} 
where $\Gamma_i$ is the greybody factor \,\cite{Page:1976df,Page:1976ki,MacGibbon:1990zk} which encodes the probability of an emitted particle surpassing the gravitational potential of a PBH and $E'$ is the total energy of the emitted particles taking into account the rotational velocity of the PBH.  

For this work, we have used {\tt BlackHawk v2.0}\,\cite{Arbey:2019mbc,Arbey:2021mbl} for the spectra of particles emitted from both non-spinning and spinning PBHs. For both non-spinning and spinning PBHs we have extensively checked the output obtained from {\tt BlackHawk}
with the analytical forms of greybody factor given in ref.\,\cite{Page:1976df}. Both primary and secondary emissions can be taken into account, though there is some uncertainty in obtaining the secondary spectrum below a certain energy range. {\tt BlackHawk v2.0} has improved on the calculation of secondary emission. We only utilized the primary emission in this work as to keep the bounds conservative. The difference in constraints with using the secondary emission is at most 60\% at the mass range $\sim 10^{15}-10^{17}$g.    
As a result of being created at a very early stage of the universe, low-mass PBHs have the potential of having a significant effect on the thermal and ionization history by means of evaporation and subsequent energy injection into the cosmic plasma.
In general, the deposited energy is related to the injected energy from Hawking emission of PBH by\,\cite{Liu:2018uzy}
\begin{eqnarray}
	\left(\frac{dE}{dVdt}\right)_{\rm dep}=f_c(z)\left(\frac{dE}{dVdt}\right)_{\rm inj},
\end{eqnarray}
where $dV$ is the volume element and $f_c(z)$ is the deposition efficiency into energy deposition channel $c$ at a given redshift $z$\,\cite{PhysRevD.94.063507,PhysRevD.95.023010,Slatyer:2015kla}. Photons, electrons and positrons emitted from a low-mass PBH heat up and ionize the IGM significantly. Here we consider energy deposition through hydrogen ionization ($f_{ion}$), helium ionization ($f_{He}$), hydrogen excitation ($f_{exc}$) and heating of IGM ($f_{heat}$). 
The energy deposited per unit volume per unit time by PBH of mass $M_{\rm PBH}$ and spin parameter $a_{*}$ is given by

\begin{eqnarray}
	\left(\frac{dE}{dVdt}\right)_{\rm dep}=\left(\frac{dE}{dVdt}\right)_{\rm dep}^\gamma + \left(\frac{dE}{dVdt}\right)_{\rm dep}^{e^\pm},
\end{eqnarray}
with
\begin{eqnarray}
		\left(\frac{dE}{dVdt}\right)_{\rm dep}^\gamma =  \int\int f_c(E_\gamma,z)E_\gamma\left(\frac{d^2N}{dE dt}\right)_\gamma \nonumber\\
		\times n_{\rm PBH}(M_{\rm PBH})\,\psi(M_{\rm PBH})\,dM_{\rm PBH}\,dE,
\end{eqnarray} 
and
\\
\begin{align}
			&\left(\frac{dE}{dVdt}\right)_{\rm dep}^{e^\pm}
			 =\int\int 2f_c(E_e-m_ec^2,z)(E_e-m_ec^2)\nonumber\\
		&\times \left(\frac{d^2N}{dE dt}\right)_{e^\pm}n_{\rm PBH}(M_{\rm PBH})\,\psi(M_{\rm PBH})\,dM_{\rm PBH}\,dE.
\end{align}
In the above equations, $n_{\rm PBH}$ is the number density of PBHs, $\psi(M_{\rm PBH})$ is the mass distribution function of the PBH. Here $n_{\rm PBH}$ can be written as
\begin{eqnarray}
	n_{\rm PBH}=f_{\rm PBH}\frac{\rho_c\Omega_{\rm DM}}{M_{\rm PBH}},
\end{eqnarray}
where $f_{\rm PBH}$ is the fraction of DM in the form of PBHs, $\rho_c$ and $\Omega_{DM}$ are the critical density and DM density parameter in the present universe.  
 {\tt DarkHistory} \,\cite{Liu:2019bbm} has implemented $f_c(z)$s that have been computed in refs.\,\cite{PhysRevD.94.063507,PhysRevD.95.023010,Slatyer:2015kla}. For our work, we have used monochromatic mass distribution, but a discussion of extended mass distribution can be followed similarly. 

The resulting modification in the temperature and ionization history due to emission from PBH can be computed from the following differential equations\,\cite{Liu:2018uzy}
\begin{eqnarray}
	\label{TPBH}
	\frac{dT_m}{dt}=\frac{dT_m^{(0)}}{dt}+\frac{dT_m^{\rm PBH}}{dt},
\end{eqnarray}
\begin{eqnarray}
	\label{xPBH}
	\frac{dx_e}{dt}=\frac{dx_e^{(0)}}{dt}+\frac{dx_e^{\rm PBH}}{dt},
\end{eqnarray}
where
\begin{eqnarray}
	\frac{dT_m^{\rm PBH}}{dt}=\frac{2 f_{heat}(z)}{3(1+f_{He}+x_{e})n_{H}}\left(\frac{dE}{dVdt}\right)_{\rm inj},
\end{eqnarray}
\\ 
\begin{eqnarray}
	\frac{dx_e^{\rm PBH}}{dt}=\left[\frac{f_{ion}(z)}{\mathcal{R}n_H}   +\frac{(1-\mathcal{C})f_{exc}(z)}{0.75\mathcal{R}n_H}       \right]\left(\frac{dE}{dVdt}\right)_{\rm inj}.
\end{eqnarray}
In the above equations $\mathcal{R}$ is the  ionization potential of hydrogen.
In presence of low-mass non-spinning and spinning PBH, the evolution of gas temperature as a function of redshift is shown in fig.\,[\ref{fig:Tm}]. As we can see in the figure, due to the presence of PBH, the IGM temperature increases significantly. The rate of emission of particles increases with the spin of the PBH. As a result, we see an increase in the gas temperature for spinning PBH with respect to the scenario with non-spinning PBH.
 \begin{figure*}[!htbp]
	\begin{center}
		\includegraphics[height=8cm]{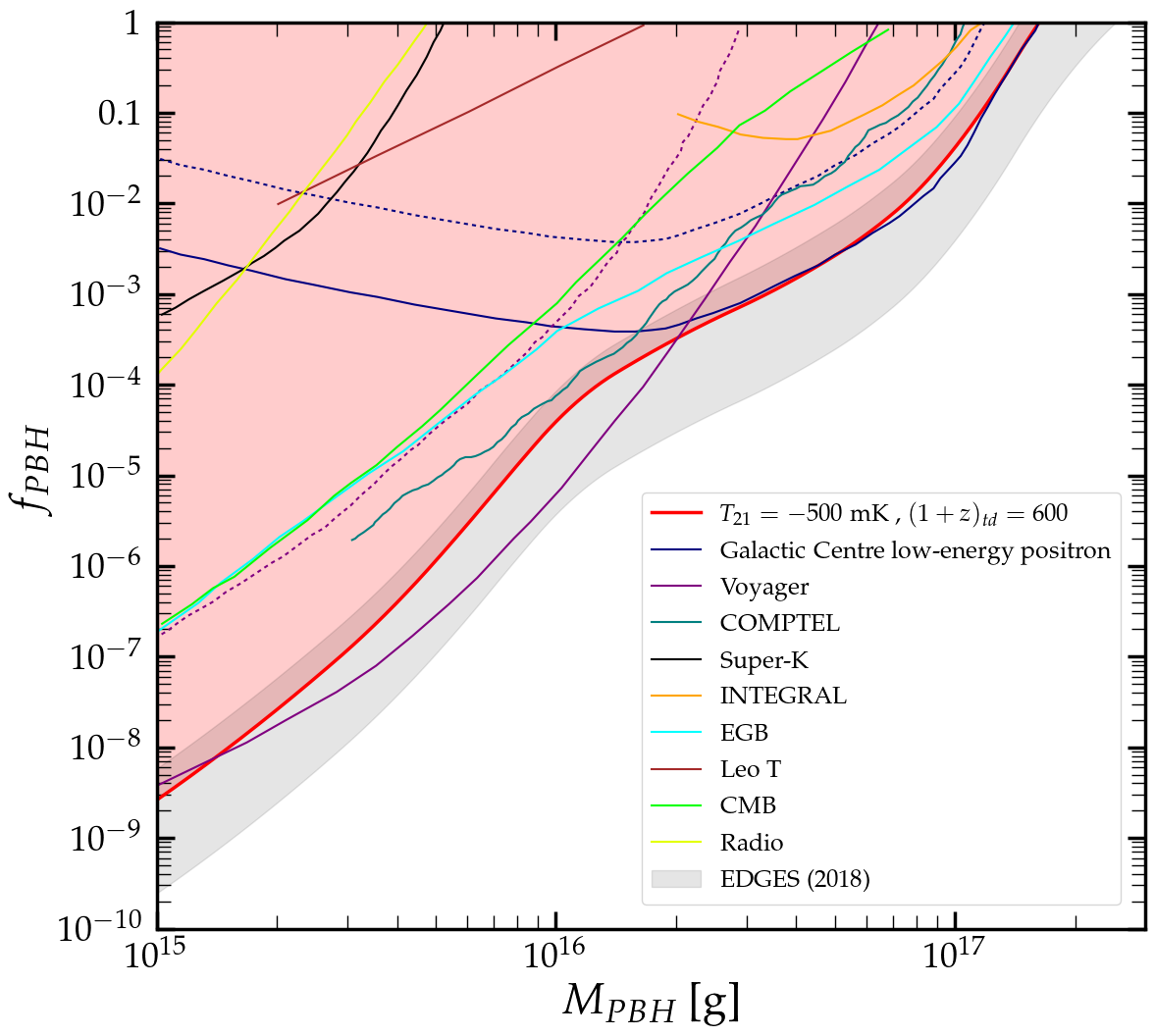}~~
		\includegraphics[height=8cm]{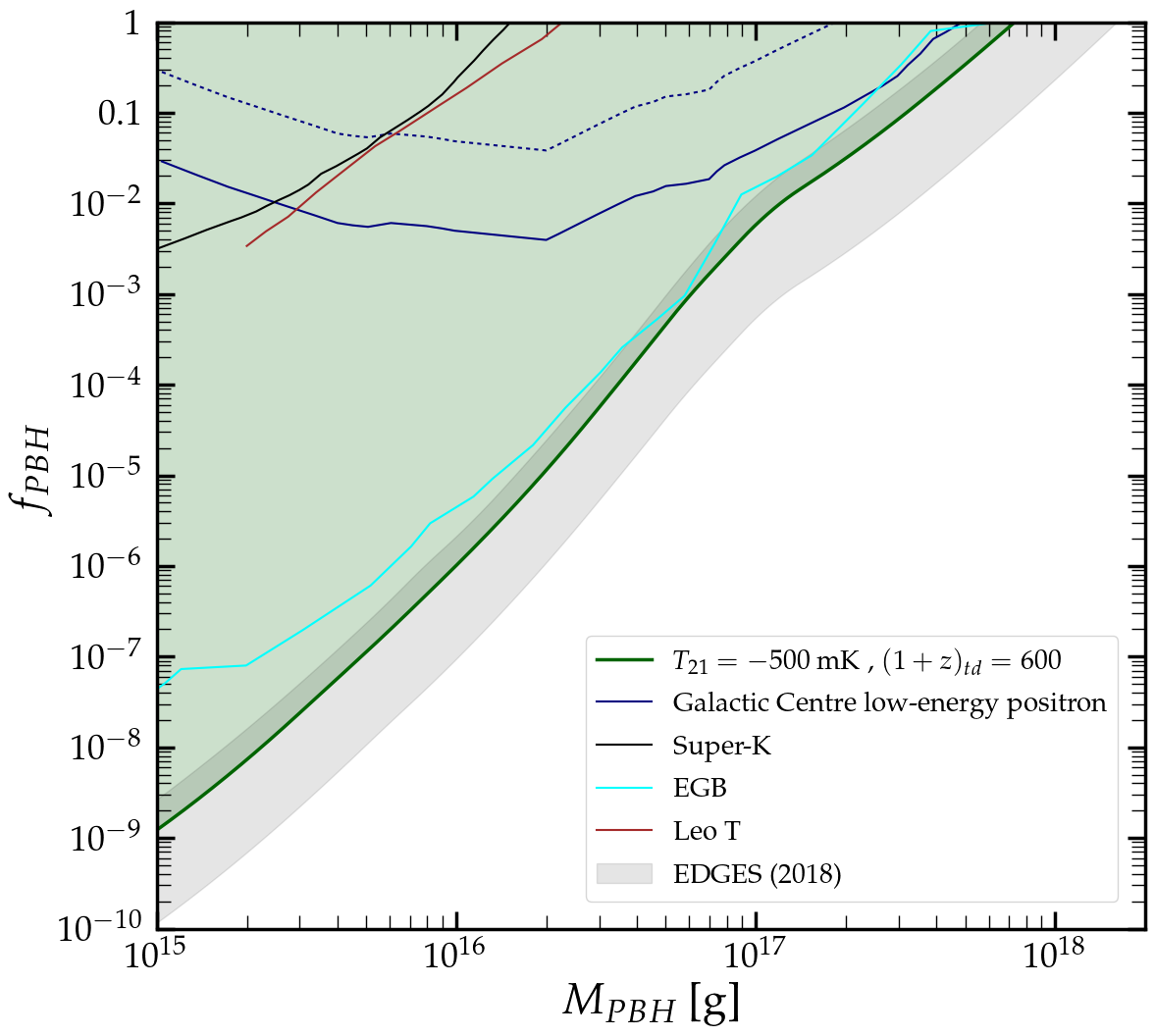}~~\\	
		\caption{Sensitivities on non-spinning ($a_*$=0) and spinning ($a_*$=0.9999) PBH DM (monochromatic distribution) from an assumed 21 cm absorption trough, similar to the one detected by EDGES  ($T_{21}=-500^{+200}_{-500}$\, mK), shown by the red (left panel) and green (right panel) lines respectively. For these limits, we assume $(1+z)_{td}=600$. The grey shaded regions show the impact of the EDGES measurement uncertainty. Previous constraints are shown as mentioned before. }
		\label{fig: -ztd}
	\end{center}	
\end{figure*}
\begin{figure}
	\centering
	\includegraphics[width=\columnwidth]{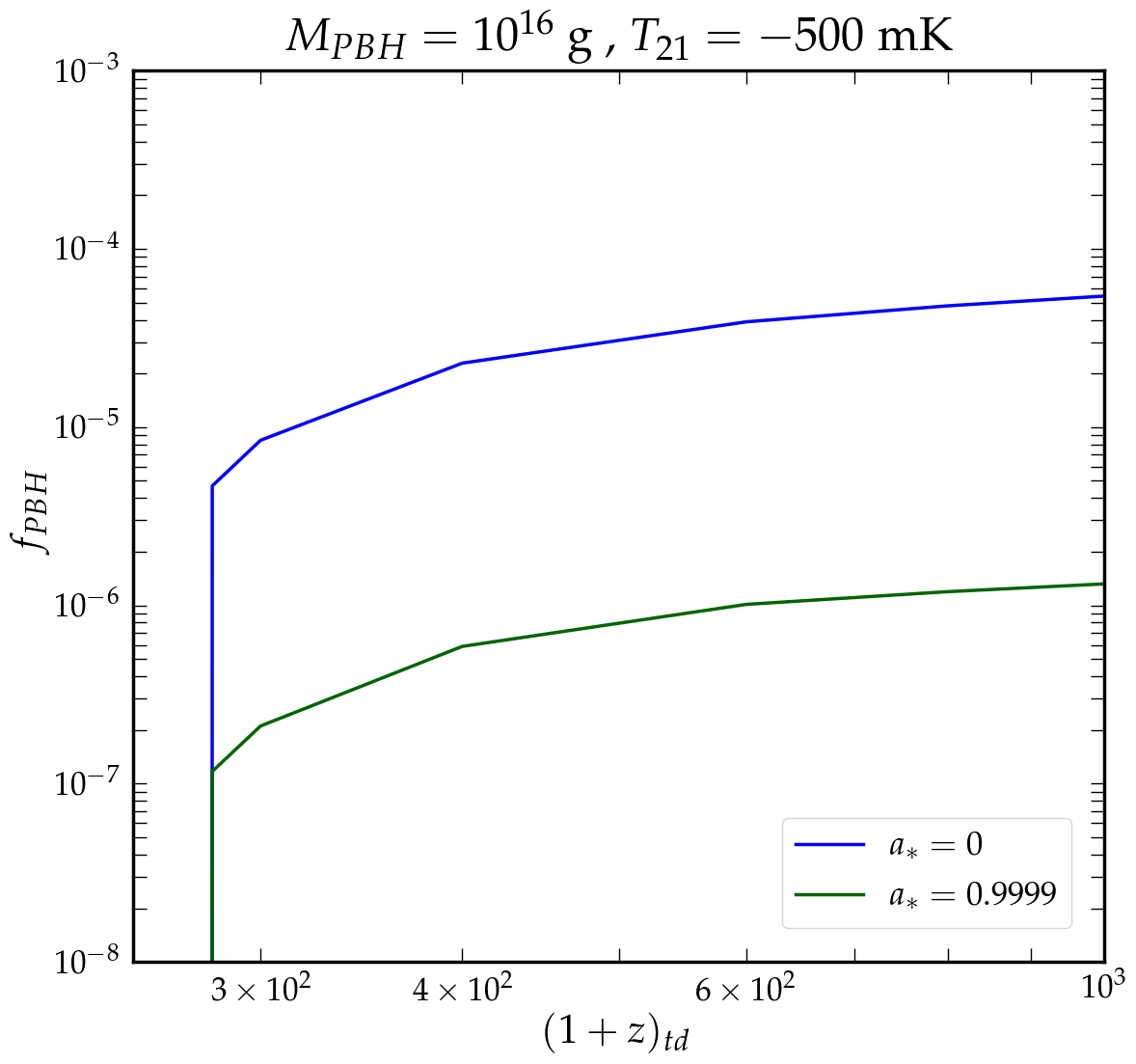}
	\caption{EDGES-like ($T_{21}=-500^{+200}_{-500}$\, mK) measurement sensitivities on both non-spinning ($a_*$=0) and spinning ($a_*$=0.9999) PBH DM (monochromatic distribution) of mass 10$^{16}$g as a function of the redshift of thermal decoupling $(1+z)_{td}$.}
	\label{fig:-Diffztd} 
\end{figure}
\section{Results}
\label{results}
In this section, we describe all the results from this work in detail. In our redshift ranges of interest, $T_m(z)\lesssim T_S(z)$. Inverting eq.\,(\ref{Brightness temp}) we can find $T_S(z)$ in terms of other variables. Combining these two relations, we get the bound on the gas temperature ($T_m$) from global 21 cm absorption troughs at a particular redshift as

\begin{eqnarray}
\label{21Bound}
    T_m(z)\lesssim T_S(z)\sim x_{HI}\left(1-\frac{0.12}{\sqrt{1+z}}T_{21}(z)\right)^{-1}T_R(z)\,\,\text{K}.\nonumber\\
\end{eqnarray}
In presence of exotic heating, we will get a particular value of $T_m(z)$ by solving eqns.\,\eqref{TPBH} and \eqref{xPBH}. Since the amount of exotic heating depends on $f_{\rm PBH}$, using the $T_m(z)$ bound obtained above, we can set upper bounds on $f_{\rm PBH}$ as a function of $M_{\rm PBH}$.
 In this work, we use the publicly available codes {\tt DarkHistory}\,\cite{Liu:2019bbm} and {\tt twentyone-global}\,\cite{caputo2020edges} to obtain our results. {\tt DarkHistory} is used for the analysis of both standard and exotic energy injection and their subsequent effect on the temperature and ionization history of IGM. We use {\tt twentyone-global} for obtaining both the standard 21 cm absorption troughs and the effect of power law radio excess on the trough at Dark Ages. 
 \begin{figure*}[!htbp]
 	\begin{center}
 		\includegraphics[height=8cm]{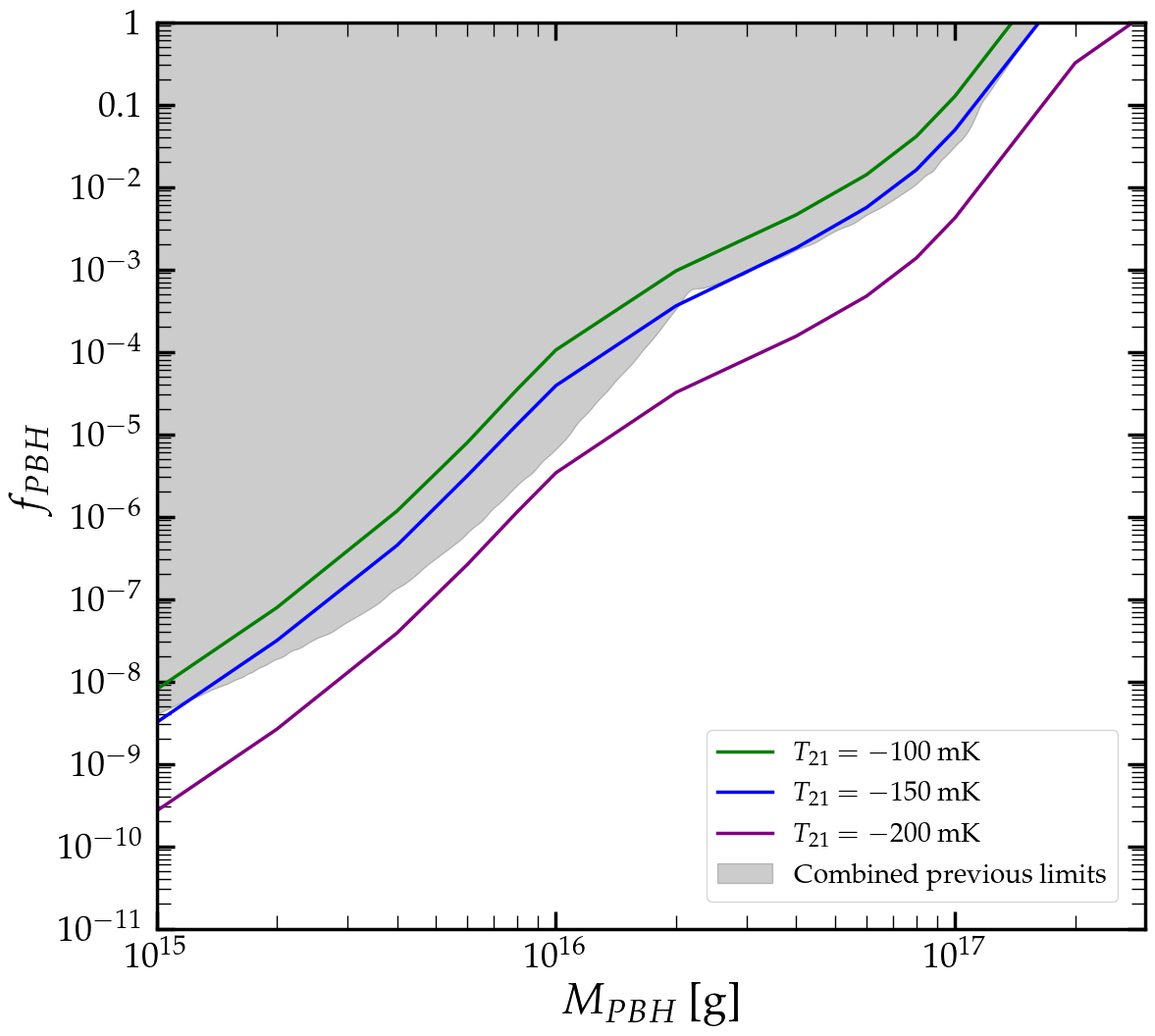}~~
 		\includegraphics[height=8cm]{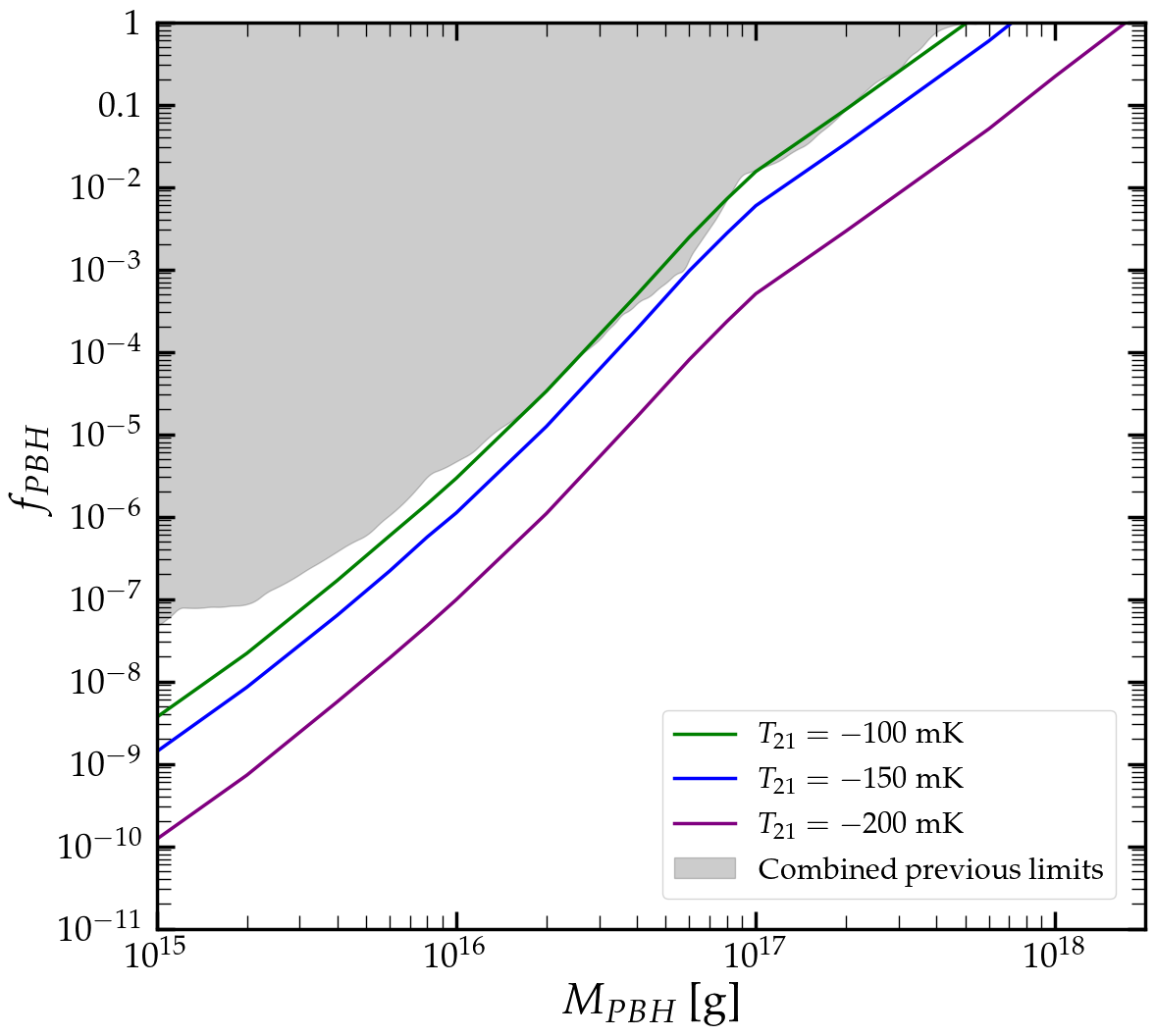}~~\\	
 		\caption{Sensitivities on non-spinning ($a_*$=0) and spinning ($a_*$=0.9999) PBH DM (monochromatic distribution) arising from the pre-reionization trough with different values of differential brightness temperature ($T_{21}$), shown in left and right panels respectively. The black shaded regions show combined previous limits as mentioned before.}
 		\label{fig: -DiffT21}
 	\end{center}	
 \end{figure*}
 
  In sections (\ref{subA}) and (\ref{secB}), we discuss the sensitivities on PBH DM coming from the pre-reionization and Dark Ages trough respectively. Here for the pre-reionization trough, we use the differential brightness temperature measured by EDGES, along with the error bars ($T_{21}= -500^{+200}_{-500}$\, mK). Standard Cosmology cannot explain an EDGES-like signal. Following the approach taken by ref.\,\cite{Liu:2018uzy}, we assume that either excess radiation at redshift\,$\sim$17.2 with respect to $T_{\rm CMB}$ or non-standard recombination  can explain an EDGES-like signal. Under these assumptions, we derive sensitivities on non-spinning ($a_*=0$) and spinning ($a_*=0.9999$) PBH DM. In section (\ref{secC}), we discuss the effect of a power law radio excess on both the absorption troughs and the respective sensitivities on PBH DM. 

\subsection{Sensitivities from the Pre-reionization Trough (EDGES-like measurement)}
\label{subA}

\subsubsection{Excess Background Radiation}
\label{sub1}
From eq.\,(\ref{Brightness temp}), we see that for a fixed value of $T_{S}$, an increase in $T_{R}$ can decrease the value of $T_{21}$. Measurements from ARCADE2 \,\cite{Fixsen_2011} and LWA1\,\cite{Dowell:2018mdb} indicate an excess diffuse radio background in the frequency ranges 3-90 GHz and 40-80 MHz respectively. If confirmed, then an excess background radiation can affect the 21 cm absorption trough and possibly explain an EDGES-like result\,\cite{Feng:2018rje,Fialkov:2019vnb}. A power law form for the excess background radiation consistent with these measurements will be discussed later.

 In this section, we remain agnostic of the origin of this excess background radiation and instead use a model independent way of considering excess radiation. We parameterize the excess radio background by taking it to be an integer multiple of the CMB temperature at the redshift of interest. Fig.\,[\ref{fig: -ExcessTR}] shows sensitivities on non-spinning ($a_*=0$) and spinning ($a_*=0.9999$) PBH DM respectively, assuming $T_{R}=5\,T_{\rm CMB}$.

  The sensitivities on spinning PBH DM are more stringent compared to non-spinning PBH DM due to the increased rate in emission of particles. In the same figures, we also show the previous constraints \,\cite{Laha:2019ssq,PhysRevLett.122.041104,PhysRevLett.126.171101,Laha:2020ivk,PhysRevLett.125.101101,PhysRevD.101.023010,Laha:2020vhg,Kim:2020ngi,PhysRevD.95.083006,Chan:2020zry} on low-mass PBH DM.
In fig.\,[\ref{fig:-DiffTR}], we show the dependence of sensitivities on different background radiation $T_{R}$, which is taken as multiples of $T_{\rm CMB}$($z=17.2$) for PBH DM with mass $10^{16}$\,g. As we increase the background radiation, the allowed upper limit on $T_m(z)$ increases. Thus, for a fixed $M_{\rm PBH}$ the allowed $f_{\rm PBH}$ increases and as a result the respective sensitivities become weaker. We notice that there is a cutoff in this plot. This is because, below a certain value of excess $T_{R}$, the value of required $T_m$ to explain EDGES-like result falls below the standard IGM gas temperature at $z\,\sim$ 17.2.

Following the discussion of fig.\,[\ref{fig:-DiffTR}], we realise that sensitivities in figs.\,[\ref{fig: -ExcessTR}] can be considerably strengthened for $2.2\,T_{\rm CMB}<T_R<5\,T_{\rm CMB}$. As a result, the sensitivities from this work can extend to some new regions of the parameter space, beyond the reach of the current constraints. We emphasize that these values of excess radiation are not actual measurements, but are taken to demonstrate the dependence of the sensitivities on the excess background radiation required to explain an EDGES-like result.

\subsubsection{Non-Standard Recombination}
\label{sec2}
During pre-reionization, $T_m\leq T_S$ and from eq.\,(\ref{Brightness temp}) we see that, a cooler than expected $T_m$ can decrease $T_{21}$ and can potentially also explain an EDGES-like signal. Various mechanisms for the exotic cooling of gas at this epoch, have been explored in the literature. Some of them are DM-baryon scattering \,\cite{Liu:2018uzy,Munoz:2015bca,Slatyer:2018aqg}, charge sequestration\,\cite{Falkowski:2018qdj} and influence of early dark energy\,\cite{Hill:2018lfx}. The last two scenarios work via altering the redshift of thermal decoupling to a larger value, implying that $T_m(z)$ evolves as $(1+z)^2$ for a longer duration, and it can possibly explain an EDGES-like excess.

 We take the approach of ref.\,\cite{Liu:2018uzy} and parameterize the early redshift of thermal decoupling as $(1+z)_{td}$. We show the sensitivities on non-spinning and spinning PBH DM from an EDGES-like signal in fig.\,[\ref{fig: -ztd}] assuming $(1+z)_{td}=600$, along with constraints from other observables.
 
  In fig.\,[\ref{fig:-Diffztd}], we show the dependence of the sensitivities on the redshift of thermal decoupling for PBH DM with mass $10^{16}$\,g.
  As we increase $(1+z)_{td}$ for a fixed mass of PBH, the gas temperature keeps on decreasing and this allows larger fraction of PBH (as DM) to be present. Thus the corresponding sensitivities become weaker. After $(1+z)_{td}$ exceeds a critical value for sufficient cooling, the dependence of sensitivities on redshift of thermal decoupling is rather weak, as can be seen in the figure. As we decrease $(1+z)_{td}$, the allowed upper limit on $T_m(z)$ decreases. Thus, for a fixed $M_{\rm PBH}$ the allowed $f_{\rm PBH}$ decreases and  the sensitivities become stronger. Here our choice of  $(1+z)_{td}=600$ and the respective sensitivities are conservative.

 If future experiments measure differential brightness temperature ($T_{21}$) in agreement with standard astrophysical expectations \,\cite{Seager:1999bc,Seager:1999km,Fialkov:2013uwm,Cohen:2016jbh,Barkana:2018lgd,Cheung:2018vww,Choudhury:2019vat}, then constraints on PBH DM can be derived without any additional assumption. In fig.\,[\ref{fig: -DiffT21}], we show the sensitivities on non-spinning and spinning PBH DM due to a range of standard $T_{21}$ (for redshift $\sim$17.2), consistent with \cite{Cohen:2016jbh}. As we decrease the value of $T_{21}$ (more negative), the bounds on gas temperature get tighter (from eq.\,\eqref{21Bound}), and with that the bounds on $f_{\rm PBH}$ get stronger.  
	 
\subsection{Projections of constraints from the Dark Ages Trough}
  \label{secB}
   \begin{figure*}[!htbp]
  	\begin{center}
  		\includegraphics[height=8cm]{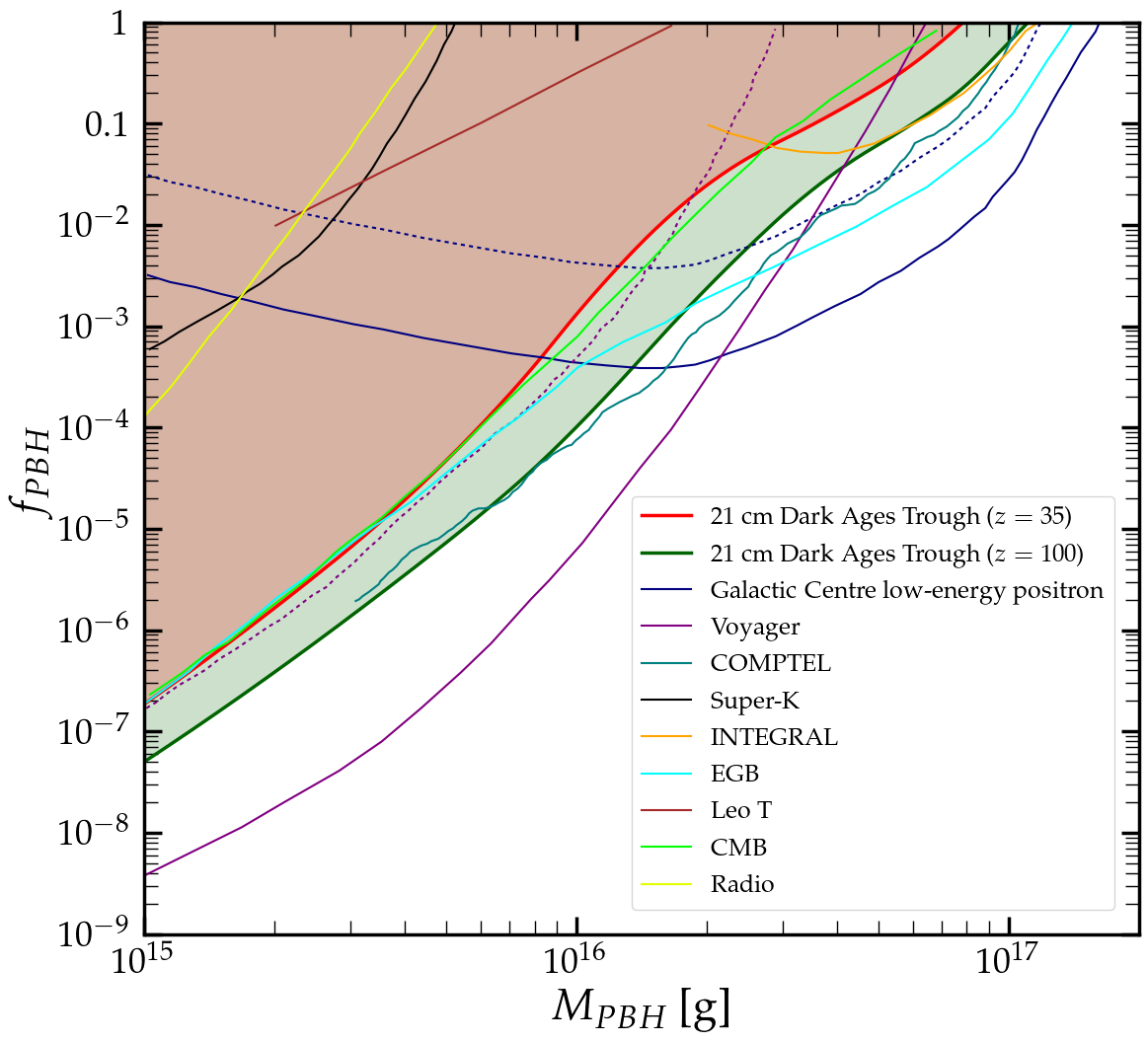}~~
  		\includegraphics[height=8cm]{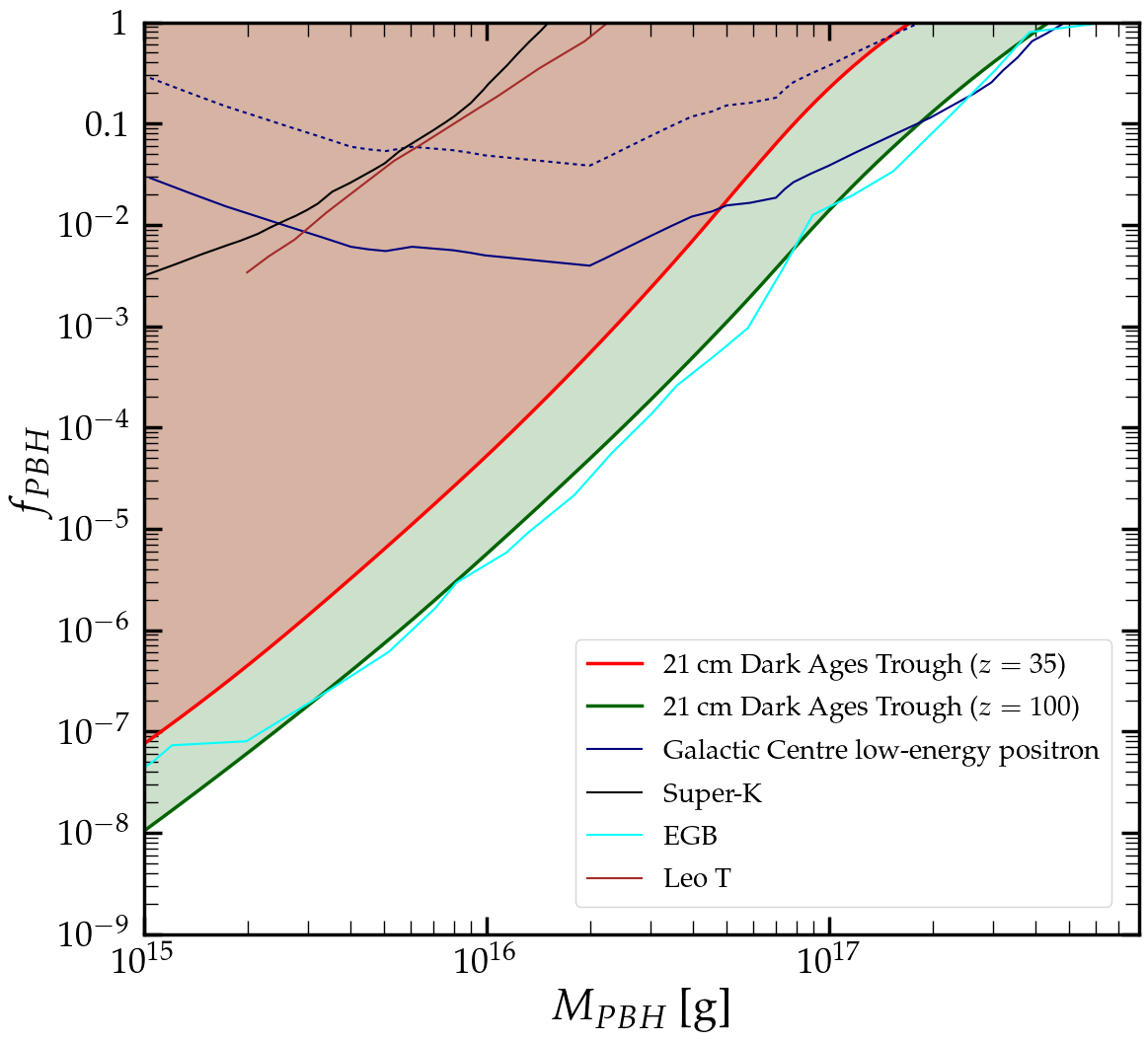}~~\\	
  		\caption{Projections of constraints on non-spinning ($a_*$=0) and spinning ($a_*$=0.9999) PBH DM (monochromatic distribution) arising from Dark Ages absorption trough ($z=35-100$), due to standard cosmology, shown in left and right panels respectively. Previous constraints are shown as mentioned before.}
  		\label{fig: -DAPPER}
  	\end{center}	
  \end{figure*}
  Similar to the pre-reionization trough, another 21 cm absorption trough is expected during the Dark Ages. Various astrophysical uncertainties do not contribute to the amplitude of this absorption trough unlike the pre-reionozation trough, though there are less severe uncertainties from the values of the cosmological parameters\,\cite{Fialkov:2019vnb}. In many cases, deviations from the standard $\Lambda$CDM predictions of the pre-reionization trough can also affect the Dark Ages trough (one such example will be discussed later). Detection of any deviation from the standard Dark Ages trough will directly point towards new physics and can potentially probe the nature of DM. Future space-based detectors like DAPPER \,\cite{Burns:2019zia}, FARSIDE\,\cite{2019BAAS...51g.178B}, DSL\,\cite{7500678}, and NCLE\,\cite{2021cosp...43E1525V} will be able to probe this signal. 
  
  For this trough, we obtain the value of standard $T_{21}(z)$ using {\tt twentyone-global}\,\cite{caputo2020edges} and then use eq.\,\eqref{21Bound} to find an upper bound on $T_m(z)$. With this upper bound, we use {\tt DarkHistory}\,\cite{Liu:2019bbm} to solve eqns.\,\eqref{TPBH} and \eqref{xPBH} to obtain the upper bound on $f_{\rm PBH}$ as a function of $M_{\rm PBH}$. We show the projections of constraints on non-spinning and spinning PBH DM from the Dark Ages trough, expected due to standard cosmology in fig.\,[\ref{fig: -DAPPER}] ($z=35-100$).
 \begin{figure}
	\centering
	\includegraphics[width=\columnwidth]{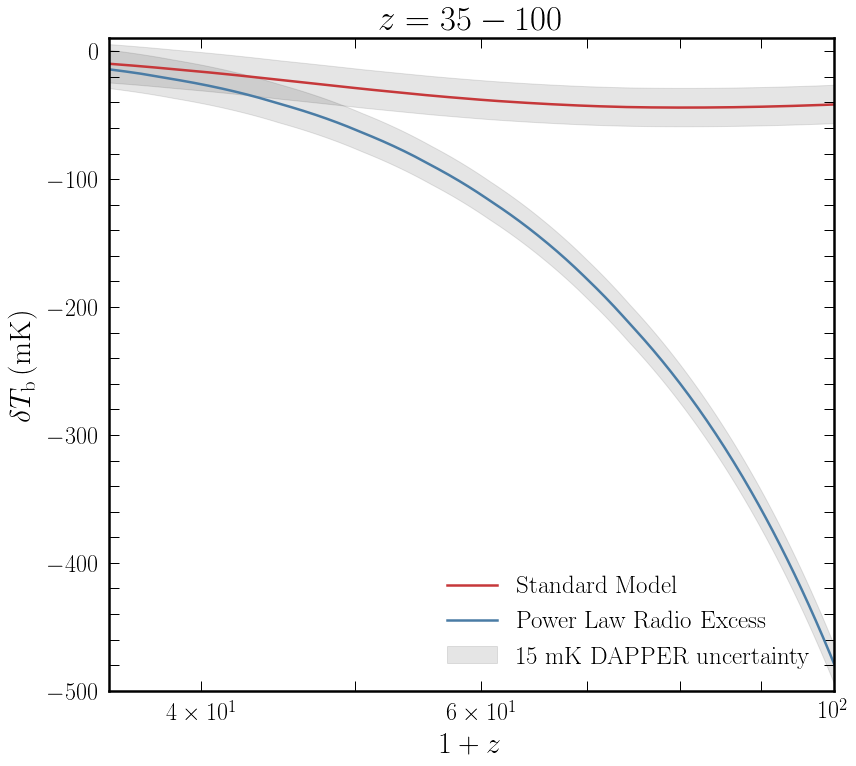}
	\caption{Differential brightness temperature expected from standard cosmology (red line) and with a power law excess background radiation with $A_r=5.7$ and $\beta=-2.6$ (blue line). The shaded region shows 15 mK DAPPER uncertainty in $T_{21}$ measurement\,\cite{Burns:2019zia}. }
	\label{fig:-DAPPER_T21} 
\end{figure}
 At higher redshifts, the increased density of PBH DM can significantly heat up the gas. Thus one expects stronger constraints from this trough. But due to the interplay between lower $T_{21}$ and higher $T_{R}$ ($T_{\rm CMB}$), one doesn't get a strong limit. These projected constraints from Dark Ages trough can be strengthened if we use measurement uncertainties assuming that the signal follows the standard $\Lambda$CDM predictions. Any constraint from the pre-reionization trough will be affected by astrophysical uncertainties unless we substantially improve both theoretical and experimental understanding of this signal. In contrast, the Dark Ages trough has much less astrophysical uncertainties, and thus is the more conservative way of probing DM in general. 
 \subsection{Effect of Power Law Radio Excess on the Absorption Troughs}
 \label{secC}
  \begin{figure*}[!htbp]
 	\begin{center}
 		\includegraphics[height=8cm]{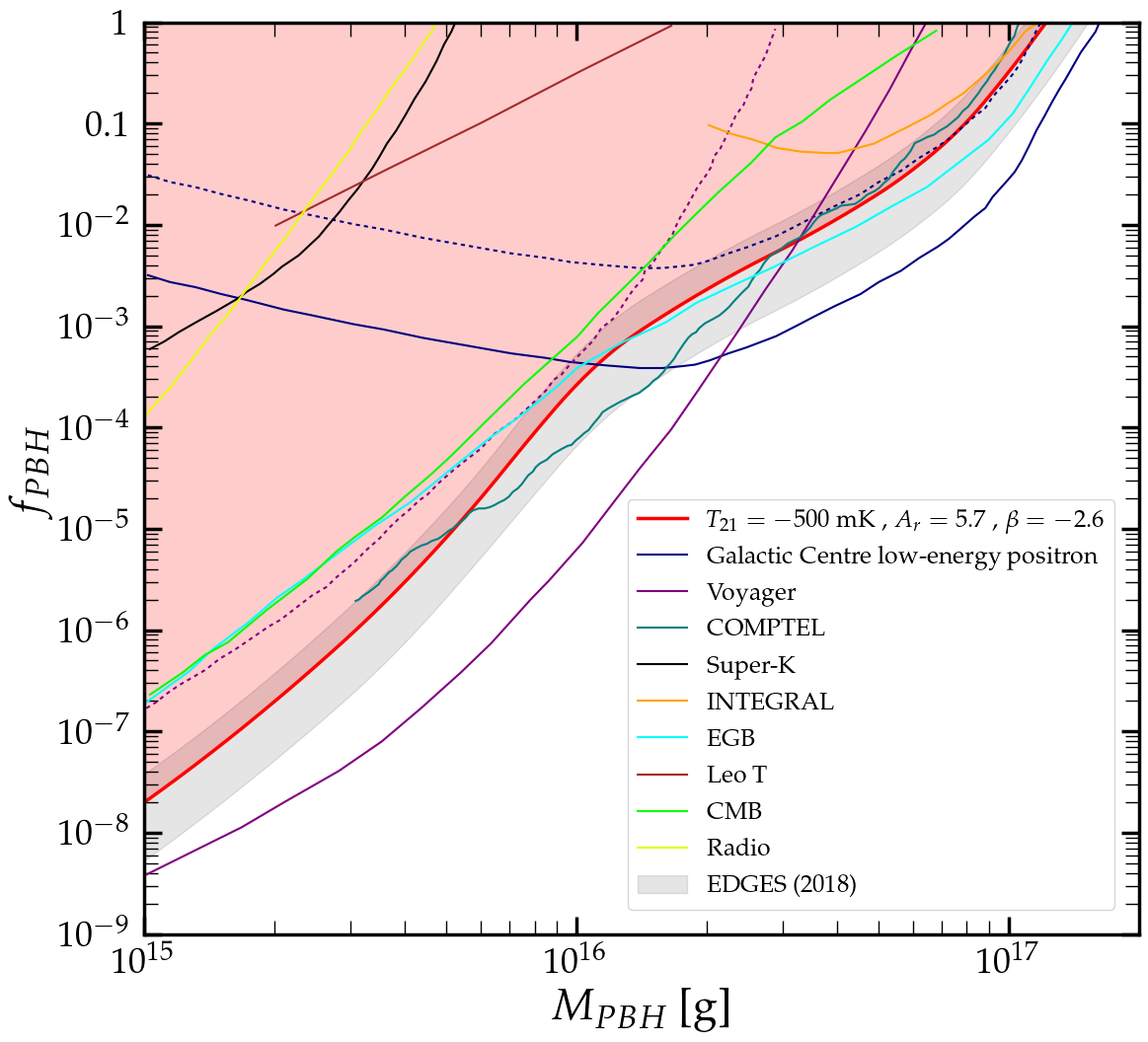}~~
 		\includegraphics[height=8cm]{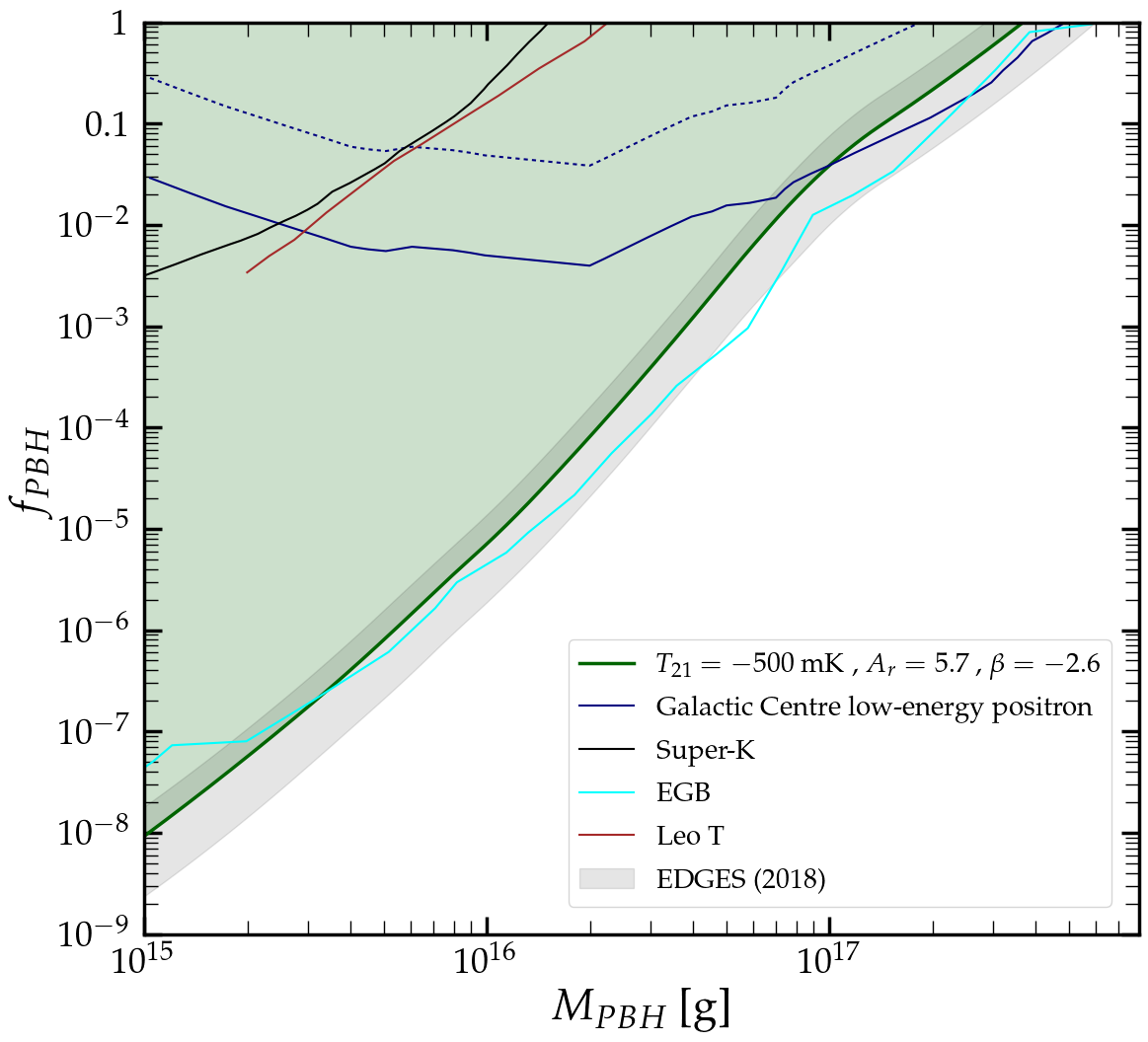}~~\\	
 		\caption{Sensitivities on non-spinning ($a_*$=0) and spinning ($a_*$=0.9999) PBH DM (monochromatic distribution) from an assumed 21 cm absorption trough, similar to the one detected by EDGES  ($T_{21}=-500^{+200}_{-500}$\, mK), shown in left and right panels respectively. Here we assume power law radio excess ($A_r=5.7$ , $\beta=-2.6$) consistent with an EDGES-like result. Previous constraints are shown as mentioned before.}
 		\label{fig: -EDGES_arcade}
 	\end{center}	
 \end{figure*}
 We have previously discussed the scenario with excess radio background and the sensitivities on PBH arising from that. Here we try to investigate the effect of a power law radio excess consistent with measurements from ARCADE2 \,\cite{Fixsen_2011} and LWA1\,\cite{Dowell:2018mdb}. The phenomenological power law is given by \,\cite{Fialkov:2019vnb}
 \begin{eqnarray}
 	\label{arcade}
 	T_{\rm rad}(z)=T_{\rm CMB,0}(1+z)\left[ 1+A_r\left(    \frac{\nu_{\rm obs}}{78\, \text{MHz}}\right)^\beta    \right],
 \end{eqnarray} 
 where $T_{\rm rad}$ is the radio background at any given redshift $z$ and $T_{\rm CMB,0}$ is the CMB temperature today (=2.7 K), $A_r$ is the amplitude defined relative to the CMB temperature, $\nu_{\rm obs}$ is the observed photon frequency and $\beta$ is the spectral
 index.
 \begin{figure*}[!htbp]
 	\begin{center}
 		\includegraphics[angle=0.0,width=0.45\textwidth]{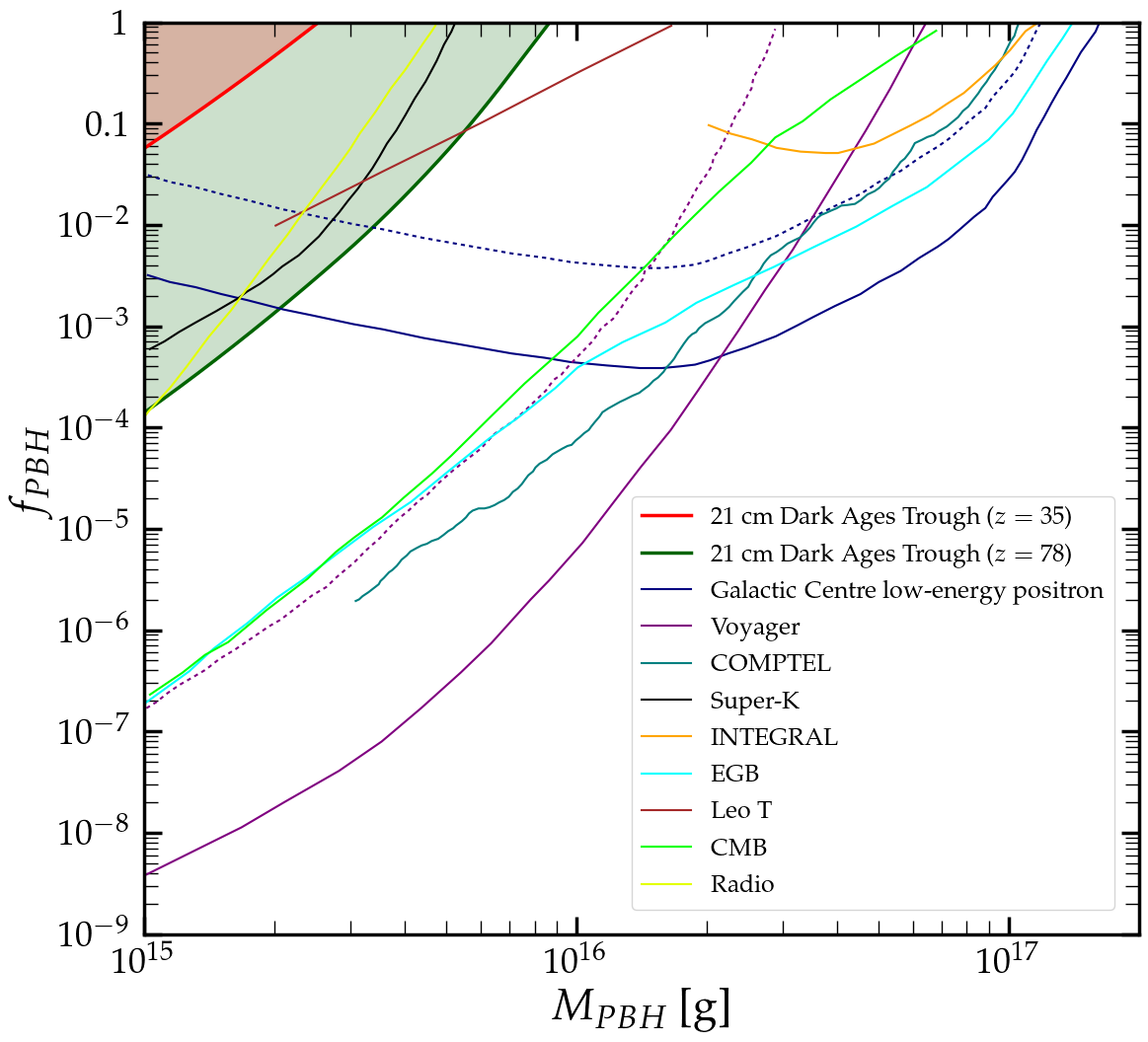}~~
 		\includegraphics[angle=0.0,width=0.45\textwidth]{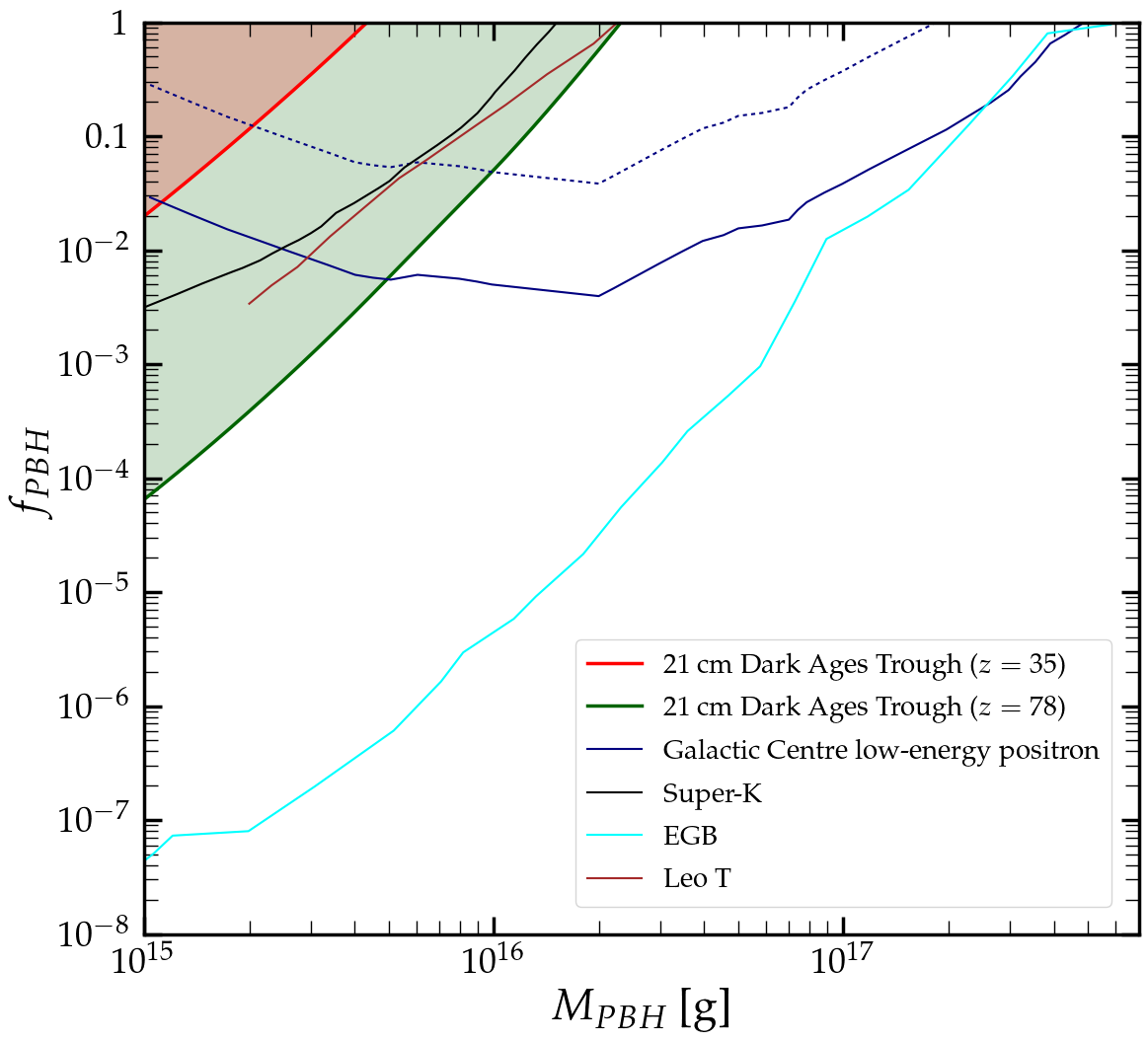}~~\\	
 		\caption{Projections of constraints on non-spinning ($a_*$=0) and spinning ($a_*$=0.9999) PBH DM (monochromatic distribution) arising from Dark Ages absorption trough ($z=35-78$), shown in left and right panels respectively.  Here we assume power law radio excess ($A_r=5.7$ , $\beta=-2.6$) consistent with an EDGES-like result. Previous constraints are shown as mentioned before. }
 		\label{fig: Dapper_Arcade}
 	\end{center}	
 \end{figure*}
 The modified radio background has significant effect on the differential brightness temperature both at pre-reionization and Dark Ages trough. This effect is shown in fig.\,[\ref{fig:-DAPPER_T21}] for the values of $A_r=5.7$ and $\beta=-2.6$ (one of the example parameters used in\,\cite{Fialkov:2019vnb}). We also show the 15 mK DAPPER uncertainty within the same plot\,\cite{Burns:2019zia}.

If we consider EDGES-like excess at the pre-reionization trough and if this occurs solely due to the power law excess radiation, then the signature must be there in the Dark Ages trough also. We use {\tt twentyone-global} to get the modified $T_{21}(z)$, assuming the radiation temperature is given by eqn.\,\eqref{arcade}. Using eq.\,\eqref{21Bound}, we get the bound on $T_m(z)$. The upper bound on $T_m(z)$ is used to obtain the bounds on $f_{\rm PBH}$ using {\tt DarkHistory}.
 
 In fig.\,[\ref{fig: -EDGES_arcade}], we show the sensitivities arising from the power law excess radiation ($A_r=5.7$ and $\beta=-2.6$) at the pre-reionization trough on both non-spinning and spinning PBH, respectively, considering an EDGES-like result ($T_{21}=-500^{+200}_{-500}$\, mK).
 Similarly in fig.\,[\ref{fig: Dapper_Arcade}], we show the projections of constraints on non-spinning and spinning PBH, respectively, considering the modified Dark Ages trough with excess power law radiation consistent with an EDGES-like excess.
 The upper limits are shown for redshift, $z=35-78$ which is within the projected redshift in which DAPPER will operate\,\cite{Burns:2019zia}. We see that at higher redshifts, the projections of constraints on both non-spinning and spinning PBHs are getting weaker, as the effect of excess background is getting stronger (see fig.\,[\ref{fig:-DAPPER_T21}]). In the left panel of fig.\,[1] from ref.\,\cite{Fialkov:2019vnb} we see that effect of different values for parameters $A_r$ and $\beta$ saturate at Dark Ages. Hence, here also the projected upper limits from Dark Ages have very weak dependence on the value of the excess radio background parameters and are thus conservative. In future, detection of the Dark Ages absorption trough will be a good test for the verification of a power law radio excess consistent with ARCADE and LWA1.
 
  All the upper limits shown are obtained without taking the effect of backreaction. The effect of backreaction changes the sensitivities between 8 to 30\%.
 Besides, there are various possible reionization models which can strengthen the sensitivities. We didn't take any particular reionization model, as a result of which all of our sensitivities are conservative.
 \section{Conclusion}
 \label{conclusion}
In this work, we derive sensitivities and projections on both non-spinning and spinning PBH DM from pre-reionization and Dark Ages absorption troughs expected in the global 21 cm signal, respectively. Recently EDGES collaboration has claimed a detection of this pre-reionization trough, though more recently SARAS 3 has rejected this detection. Our work explores the bounds expected from any future detection of global 21 cm excess similar to that of EDGES. For pre-reionization trough, we use the EDGES signal as the benchmark and derive conservative sensitivities assuming two scenarios, i.e., excess background radiation (at $z \sim$ 17.2) and non-standard recombination. These limits are comparable with existing limits. Our limits can be significantly strengthened by taking slightly lower value of excess radiation temperature and higher redshift of thermal decoupling. We also consider the scenario where future measurement of global 21 cm pre-reionization trough yields a differential brightness temperature that is within the standard astrophysical expectation. Using that we derive the corresponding constraints on PBH DM and notice that for lower values of $T_{21}$ we can probe new parts of PBH DM parameter space. For the Dark Ages trough, we use the absorption trough expected from standard $\Lambda$CDM model to obtain our projections. Later we discuss the possibility of a power law excess radiation hinted by both ARCADE and LWA1 measurements. We derive sensitivities on $f_{\rm PBH}$ from an EDGES-like signal and Dark Ages trough in the presence of such excess power-law background radiation. The Dark Ages trough is still beyond the reach of current global 21 cm signal detectors, but in future, space based telescopes will be able to detect such signal. A combined detection of
both the absorption troughs will in future provide the best
insight to the Universe during both Dark Ages and Cosmic
Dawn, and using these measurements one can either discover or constrain various well motivated beyond Standard Model physics scenarios. In addition to various near-future 21 cm experiments, low-mass PBH DM can also be discovered by various upcoming neutrino, low-energy gamma-ray, cosmic ray, CMB, radio experiments and measurements sensitive to Galactic Centre low-energy positrons\,\cite{PhysRevLett.126.171101,DeGouvea:2020ang,Dutta:2020lqc,Wang:2020uvi,Cang:2020aoo,Ray:2021mxu,DeRomeri:2021xgy,Ghosh:2021gfa,Capanema:2021hnm,Mukhopadhyay:2021puu}. A combination of all of these experiments will fully explore the low-mass PBH DM parameter space.
 
 Note Added: Refs.\,\cite{Mittal:2021egv,Natwariya:2021xki,Cang:2021owu} appeared in arXiv while this work was in preparation.
\section{Acknowledgement}
We especially thank Hongwan Liu for detailed help regarding {\tt DarkHistory}, and also for his comments on the manuscript. We also thank Julian B.\, Muñoz, Tarak Nath Maity and Shikhar Mittal for their useful comments and suggestions. R. L. acknowledges financial support from the
Infosys foundation, Bangalore and institute start-up funds. 
\newpage
\bibliographystyle{JHEP}
\bibliography{ref.bib}

\end{document}